\documentclass{aa}  
\pdfoutput=1
\usepackage{graphicx}
\usepackage{subfigure}
\usepackage{txfonts}
\usepackage{booktabs}
\usepackage{blindtext}
\usepackage[draft]{hyperref}

\begin{document}

  \title{Ly$\alpha$-Lyman Continuum connection in $3.5\leq z\leq 4.3$ star-forming galaxies from the VUDS survey
                             \thanks{Based on data obtained with the European 
                                       Southern Observatory Very Large Telescope, Paranal, Chile, under Large
                                       Program 185.A--0791. }
                  }

            \author{F. ~Marchi\inst{1}
            	\and L.~Pentericci\inst{1}
            	\and L.~Guaita\inst{13}
            	\and D.~Schaerer\inst{3,4}
            	\and A.~Verhamme\inst{3}
            	\and M.~Castellano\inst{1}
            	\and B.~Ribeiro\inst{2}
            	\and B.~Garilli\inst{5}
            	\and O.~Le F\`evre\inst{2}
            	\and R.~Amorin\inst{11,12}
            	\and S.~Bardelli\inst{8}
            	\and P.~Cassata\inst{7}
            	\and A.~Durkalec\inst{10}
            	\and A.~Grazian\inst{1}
            	\and N.P.~Hathi\inst{6}
            	\and B.~C.~Lemaux \inst{9}
            	\and D.~Maccagni\inst{5}
            	\and E.~Vanzella\inst{8}
            	\and E.~Zucca \inst{8}
            }       
            
            \institute{INAF–Osservatorio Astronomico di Roma, via Frascati 33, 00040 Monte Porzio Catone, Italy\\
            	\email{francesca.marchi@oa-roma.inaf.it}
            	\and
            	Aix Marseille Universit\'e, CNRS, LAM (Laboratoire d'Astrophysique de
            	Marseille) UMR 7326, 13388, Marseille, France
            	\and
            	Geneva Observatory, University of Geneva, ch. des Maillettes 51, CH-1290 Versoix,
            	Switzerland
            	\and
            	Institut de Recherche en Astrophysique et Plan\'etologie - IRAP, CNRS, Universit\'e
            	de Toulouse, UPS-OMP, 14, avenue E. Belin, F31400
            	Toulouse, France  
            	\and 	 
            	INAF--IASF Milano, via Bassini 15, I--20133, Milano, Italy                               
            	\and
            	Space Telescope Science Institute, 3700 San Martin Drive, Baltimore, MD 21218, USA
            	\and
            	Instituto de Fisica y Astronom\'ia, Facultad de Ciencias, Universidad de
            	Valpara\'iso, Gran Breta\~{n}a 1111, Playa Ancha, Valpara\'iso Chile
            	\and
            	INAF--Osservatorio Astronomico di Bologna, Via Gobetti 93/3 - 40129, Bologna, Italy
            	\and
            	Department of Physics, University of California, Davis, One Shields Ave., Davis, CA 95616, USA
            	\and
            	National Center for Nuclear Research, ul. Hoza 69, 00-681, Warszawa, Poland
             	\and 
                Cavendish Laboratory, University of Cambridge,
            	19 JJ Thomson Avenue, Cambridge, CB3 0HE, UK    
            	 \and
            	 Kavli Institute for Cosmology, University of Cambridge,
                Madingley Road, Cambridge CB3 0HA, UK
                \and
                Nùcleo  de  Astronomìa,  Facultad  de  Ingenierìa,  Universidad  Diego  Portales,  Av.    Ejèrcito  441,  Santiago,  Chile
            }
        
           \date{Received; accepted}

   \date{}
    \abstract
    {To identify the galaxies responsible for
    	the reionization of the Universe,  we must rely on the investigation of the Lyman Continuum (LyC) properties of $z\lesssim 5$ star-forming galaxies, where we can still directly observe  their ionizing radiation.}
    {The aim of this work is to explore  the correlation between the LyC
    	emission and some of the proposed indirect indicators of LyC radiation at $z\sim
    	4$ such as a bright Ly$\alpha$ emission and a compact UV continuum size.}
    {We selected a sample of 201 star-forming galaxies from the Vimos
    	Ultra Deep Survey (VUDS) at $3.5\leq z \leq 4.3$ in the COSMOS,
    	ECDFS and VVDS-2h fields, including only those with reliable
    	spectroscopic redshifts, a clean spectrum in the LyC range and
    	clearly not contaminated by bright nearby sources in the same
    	slit. For all galaxies we have measured the Ly$\alpha$ EW, the Ly$\alpha$ velocity shift with respect to
    	the systemic redshift, the Ly$\alpha$ spatial extension and the UV continuum
    	effective radius.  We then selected different sub-samples
    	according to the properties predicted to be good LyC emission
    	indicators: in particular we created sub-samples of galaxies with $EW(Ly\alpha)\geq70\,\AA$, $Ly\alpha_{ext}\le5.7\,kpc$, $r_{UV}\le0.30\,kpc$ and  $|\Delta  v_{Ly\alpha}|\leq 200\,km/s$.
    	We stacked all the galaxies in each sub-sample and measured the flux
    	density ratio ($f_{\lambda}(895)/f_{\lambda}(1470)$), that we
    	consider to be a proxy for LyC emission, and compared these ratios to
    	those obtained for the complementary samples.
    	Finally, to estimate the statistical contamination from lower redshift 	interlopers in our samples, we performed dedicated Monte Carlo simulations using   an ultradeep U-band image    	of the ECDFS field.}
    {We find that the stacks of galaxies which are  UV compact ($r_{UV}\le0.30\,kpc$) and
    	have bright  Ly$\alpha$ emission ($EW(Ly\alpha)\geq70\,\AA$), have much higher LyC fluxes
    	compared to the rest of the galaxy population. These parameters  appear to be good indicators of LyC radiation in agreement with theoretical studies and previous
    	observational works.
    In addition we find that galaxies with a low Ly$\alpha$ spatial extent ($Ly\alpha_{ext}\le5.7\,kpc$) 	have higher LyC flux compared to the rest of the population: such a correlation had never been
    	analysed before and seems even stronger than the correlation with     	high EW(Ly$\alpha$) and small $r_{UV}$. These results assume that the stacks from all the sub-samples 
    	present the same statistical contamination from lower redshift interlopers.
    	If we subtract a statistical contamination from low redshift interlopers obtained with the simulations from the flux density ratios ($f_{\lambda}(895)/f_{\lambda}(1470)$) of the significant sub-samples we find that these samples contain real LyC leaking flux with a very high probability, although the true average escape fractions are very uncertain.}  
    {Our work indicates that galaxies with very high $EW(Ly\alpha)$, small  $Ly\alpha_{ext}$ and small  $r_{UV}$ are very likely the best candidates to show Lyman Continuum radiation at $z\sim 4$ and could therefore be the galaxies that contributed more to reionization.} 
    
        \titlerunning{}
        \authorrunning{Marchi F.}
   \maketitle        
   
\section{Introduction}
        \label{sec:intro}
        Understanding the processes that led to the reionization of the Universe is among the most challenging tasks of modern extra-galactic  astronomy. The most likely objects responsible for this phenomenon were star-forming galaxies and active galactic nuclei (AGN) that, at $z\sim6$, completely ionised the intergalactic medium (IGM) thanks to the emission of the so-called Lyman continuum (LyC) radiation \citep{robertson10,shull12,robertson15,becker13,giallongo15}, which is at $\lambda <912\,\AA$. However, at  redshift  higher than $z\sim4.5,$ the IGM becomes less transparent to LyC photons due to the increasing number of
        intervening absorption systems and can prevent the direct detection of Lyman continuum flux \citep{madau95,madau99,dijkstra04,inoue08,prochaska09,laursen11,inoue:transmissivity,worseck14}. It is therefore not possible to directly study the LyC emission of the sources responsible for the reionization. What we can do, is to study the ionising emission properties of lower redshift galaxies  and later infer if these properties are more common during the reionization epoch. 
        
        At $z \gtrsim 2.5$, the LyC radiation is redshifted into the optical spectral
        region and therefore, for galaxies in the range $2.5 \lesssim z \lesssim 4.5$, it can be detected using ground-based observations. For galaxies at lower redshift, we must instead rely
               on space-based observations. The ionizing radiation is significantly attenuated by neutral gas and dust in the interstellar and circumgalactic
               medium of the sources \citep{leithere95,deharveng01}. Therefore, the detection of LyC emission
               in individual galaxies is a rather difficult task. Furthermore, at high redshift the search for LyC emitters (LCEs) is made much complicated by the high probability of contamination by lower redshift interlopers, the faintness of the sources and the increase of the IGM opacity with redshift \citep{vanzella10,vanzella12,inoue:transmissivity}.
              In particular, the line
               of sight (LoS) contamination is one of the main limitations of LyC
               studies when imaging and spectroscopic observations are taken
               from the ground \citep{vanzella12}. Indeed low-redshift
               galaxies can mimic the LyC emission from high-redshift sources
               if they are located very close to the target galaxies and the
               spatial resolution does not allow to distinguish the two objects. These
               nearby contaminants can only be identified in high-resolution
               HST images because they appear blended in ground-based observations.
               In most cases, the putative LyC emission appears offset in HST images with respect to the main optical galaxy, indicating the presence of a possible lower-redshift contaminant \citep{nestor,mostardi,grazian16}.
               
               Blind searches for LCEs have not been, indeed, very productive so far. Only three LyC emitters have been found with blind searches in the local Universe \citep{bergvall06,leitet11,leitet13,leitherer16} and only two detections have been reported at high redshift \citep{shapley16,mostardi15}.
 To overcome this lack of detections, several pre-selection methods to find good LyC leaker candidates have been proposed, leading to the discovery of six further LCEs at low redshift \citep{borthakur14,izotov16} and one LCE at high redshift \citep{vanzella16}. It has been found indeed that some galaxy properties can be related to a high escape fraction of ionizing radiation. These features are the
        nebular emission line strengths \citep{zackrisson13}, high
        $[OIII]\lambda  5007/[OII]\lambda3727$ ratios that could trace density-bounded HII regions
        \citep{jaskot,nakajima14} and the non-saturation
        of the metallic low-ionisation absorption lines that trace
        a low covering fraction of the absorbing gas along the line of
        sight \citep{heckman11,alexandroff,vasei16}. It is also believed that the faintest, low-mass star forming galaxies  are 
        responsible for the bulk of the ionizing radiation during the reionization epoch: 
     the    observed UV luminosity function is very steep at high-z and these faint galaxies should be very numerous  \citep{ouchi09,fesc2,Yajima11,bouwens12,mitra13,mason15} hence providing the necessary ionizing budget.
        Since high redshift Lyman Alpha Emitters (LAEs) are in general low-mass galaxies \citep[e.g.][]{finkelstein07,bouwens07,pentericci10}, characterized by a faint UV continuum, a bright Ly$\alpha$ emission line has also been proposed as a pre-selection tool to look for LyC emitters \citep{verhamme15,dijkstra16}.  UV morphology can also be used as an indirect indicator of LyC emission: very compact star-forming regions can photoionize the ISM creating the so-called density-bounded regions \citep{nakajima14} or can shape low density channels through the ISM by mechanical feedback, facilitating the escape of LyC and  Ly$\alpha$ radiation. A connection between the LyC, Ly$\alpha$ emissions and UV compactness is therefore expected at some level. 

        There have been several theoretical studies investigating the above correlations. For example \cite{dijkstra16} explore the correlation between LyC and Ly$\alpha$ radiation using a large suite of simplified models of the multi-phase ISM that span the wide range of astrophysical conditions encountered in observed galaxies. They model the source of LyC and Ly$\alpha$ radiation surrounded by a collection of spherical clumps of dust and neutral hydrogen gas which are opaque to LyC radiation and that are embedded within a hot inter-clump medium \citep[as in][]{hansen06,laursen13,gronke14}. In these models LyC photons can escape from the galaxy if their sight-line does not encounter any clump. They find that galaxies with a low escape fraction of  Ly$\alpha$ photons ($f_{esc}^{Ly\alpha}$), also present a low escape fraction of LyC photons ($f_{esc}^{LyC}$), while galaxies with high values of $f_{esc}^{Ly\alpha}$ present a large spread in $f_{esc}^{LyC}$. 
        Finally, they find that galaxies that show LyC emission typically have narrower and more symmetric Ly$\alpha$ line profiles and a low velocity offset with respect to the systemic redshift.

        \cite{verhamme15} use a similar approach to study the LyC-Ly$\alpha$ connection. They use the classic shell model to picture the galaxy in two different configurations: 1) totally ionized ISM and 2) riddled (i.e. with ionised holes) ionization-bounded ISM. In the first case they find a Ly$\alpha$ spectrum characterized by a very narrow profile with a small shift with respect to the systemic redshift \citep[as][]{dijkstra16} whereas in the second case they find a Ly$\alpha$ peak well centered at the systemic z but with a flux redwards. They find also that, for  galaxies with very low or null outflow velocity for which the Ly$\alpha$ profile is characterized by a double peak, a small separation between the two peaks is a strong indicator of LyC emission.


       As mentioned before, these indirect indicators have been partly confirmed by observations.
     \cite{izotov16} showed that a selection
    for compact star-forming galaxies showing high $[OIII]\lambda  5007/[OII]\lambda3727$ ratios (>5) appears
    to pick up very efficiently sources with escaping Lyman continuum radiation at low redshift: they find LyC emission from all the five galaxies selected by these criteria, considerably increasing the number of known LCEs at low redshift. On the other hand, \cite{rutkowski17}, selecting a sample of $z\sim2$ star-forming galaxies with the same constraint on the $[OIII]\lambda  5007/[OII]\lambda3727$ ratio as \cite{izotov16}, did not find any individual detections. 
    
     \cite{verhamme17} analyzed the Ly$\alpha$ spectral properties of several known LyC emitters finding that all the LCEs in the local Universe are characterized by a double peak Ly$\alpha$ profile with a small peak separation in agreement with the theoretical expectations. Furthermore, they find that several known LyC sources present a Ly$\alpha$ in emission with very high EW  and a large Ly$\alpha$ escape fraction ($f_{esc}^{Ly\alpha}>20\%$) as predicted by \cite{dijkstra16}.  Finally, they observe a correlation between the escape fraction of ionizing photons and the SFR surface density. This is an evidence that the compactness of star-forming regions could play a significant role in the escape of ionizing radiation.

        In our previous paper \citep{marchi}, we also found an indication of a possible positive trend of the flux density ratio ($f_{\nu}(910)/f_{\nu}(1500)$) as a function of Ly$\alpha$ EW using a sample of 12 LAEs at $z\sim 3.8$ in the \emph{VIMOS Ultra Deep Survey} \citep[VUDS,][]{lefevreVIMOS}. A similar correlation has been also found by \cite{micheva15} with a sample of 18 LAEs at $z\geq 3.06$ in the SSA22 proto-cluster. These studies give support to the correlation between the  escape of Ly$\alpha$ photons (that is correlated to the Ly$\alpha$ EW) and the escape of LyC radiation. 

Finally we mention \emph{Ion2}, a high redshift galaxy that was initially identified as a possible LyC emitter by \cite{vanzella15}: subsequent observations confirmed the presence of LyC radiation from this source \citep{vanzella16,debarros17}. This galaxy also presents many physical properties typical of LCEs and discussed above as indirect indicators (e.g. small UV size and high Ly$\alpha$ EW). Finally, six more LyC emitting candidates,  have been recently proposed by \cite{naidu16} but their LyC emission has still to be verified.

    These latest results seem to indicate that the use of indirect indicators to identify the best LyC emitters candidates could be very efficient. In this paper we extend this analysis to high redshift to test some of the proposed trends, using a large sample of galaxies at $z\sim4$ with spectra from the VUDS survey.
         Thanks to this large dataset, we can select different sub-samples of galaxies according to the properties predicted to be good LyC emission indicators and then test if the population of galaxies in each sub-sample does present an excess in the LyC part of the spectrum. 
         
         This paper is organised as follows. In Section \ref{sec:samplesel} we describe the selection criteria that we used to select our sample of high redshift star-forming galaxies from the VUDS dataset. In Section \ref{sec:method} we describe the methods to evaluate the different galaxies' properties, the criteria to select each sub-sample and the technique to evaluate the flux density ratio,  $\frac{f_{\lambda}(895)}{f_{\lambda}(1470)}$, for each sub-sample. In Section \ref{results} we present the obtained results. Finally, in Section \ref{sec:contamination} we discuss the effects of contamination from lower redshift interlopers on our samples.
          Throughout the paper we adopt the $\Lambda$ cold dark matter ($\Lambda$-CDM) cosmological model ($H_0 = 70\, \mathrm{km\, s^{-1} Mpc^{-1}}$, $\Omega_M = 0.3$ and $\Omega_{\Lambda} = 0.7$).
          All magnitudes are in the AB system. All EWs presented in this paper are given in the rest-frame and positive values correspond to those lines measured in emission.
         
        \section{Sample selection}
        \label{sec:samplesel}
                               \begin{figure*}[h]
                               	\centering
                               	\includegraphics[width=0.6\textwidth]{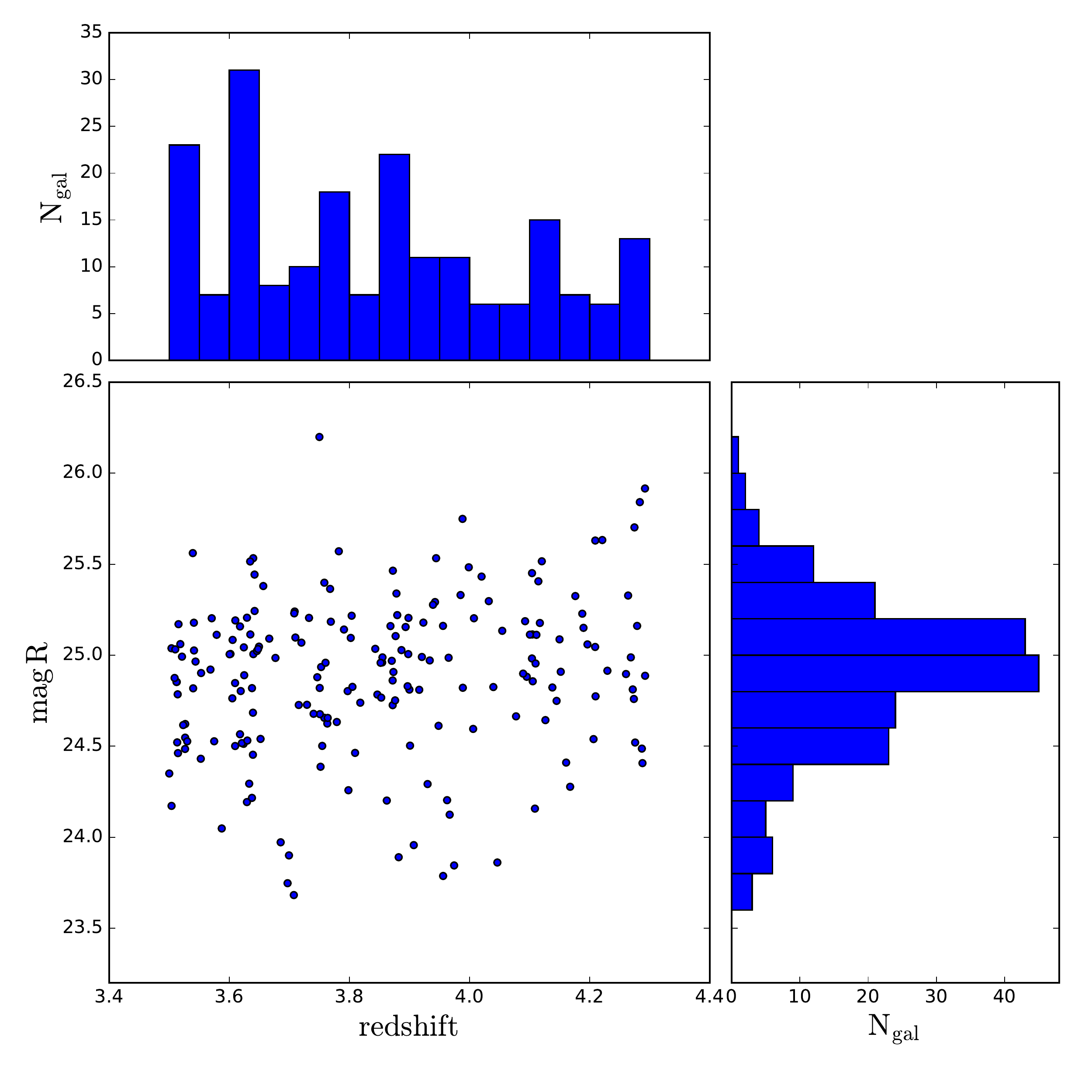}
                               	\caption{R magnitude as a function of the redshift for the 201 star-forming galaxies in the final sample. The distributions of the two quantities are shown on the sides of the plot.}
                               	\label{fig:histoz}
                               \end{figure*}
             
               We selected a sample of star-forming galaxies from the VUDS
                database \footnote{The first  public release is available at \href{url}{http://cesam.lam.fr/vuds/DR1/} but our sources  are selected from the entire database} \citep{lefevreVIMOS,tasca17}, which is the largest, up to date,  spectroscopic survey of galaxies at $z>2$. VUDS acquired approximately $7000$ spectra of galaxies at $2\leq z\leq 6$ in the COSMOS, ECDFS and VVDS-2h fields. We selected all galaxies with reliable redshifts in the range $3.5 \leq z \leq 4.3$. In order to measure a possible LyC
               signal in the wavelength range covered by the VUDS spectra ($3800-9400\, \AA$), we require sources at $z > 3.5$  to observe the LyC wavelength domain redshifted into the spectral interval of the VUDS observations.
               The upper redshift limit is related to the almost totally opaque IGM at $z > 4.5$ \citep{madau95,laursen11,inoue:transmissivity}.  We however use a more restricted cut at  $z=4.3$ to maximise the Signal to Noise (given that at $z>4.3$ there are only few good quality spectra).
               
               The redshift determination of the VUDS targets is explained in details in \cite{lefevreVIMOS}. Briefly,  each spectrum is analysed by two different VUDS team members using the EZ tool \citep{garilli10}, a cross-correlation
               engine to compare spectra and a wide library of galaxy and star
               templates, along with a visual inspection of the spectra when only emission lines are present.
               A quality flag is assigned to each redshift, following the scheme that was tested in previous surveys \citep[e.g. VVDS,][]{lefevre14}.
  In our analysis we included only galaxies with VUDS reliability flags 3, 4, 23 and 24 corresponding to a probability greater than $95 \%$ for the spectroscopic redshift to be correct \citep[see][for more details]{lefevreVIMOS}.  The spectra are calibrated using spectrophotometric standard stars with a relative flux accuracy of better than $\sim$5\% over the wavelength range $3600$ to $9300\,\AA$. In addition, each spectrum is corrected for atmospheric extinction and for wavelength-dependent slit losses due to atmospheric refraction, taking into account the geometry of each source as projected into the slit \citep{thomas17}. The spectra are also corrected for the galactic extinction \citep{lefevreVIMOS}.
 With an integration of 14 net hours per target per grism, the VUDS spectra reach a S/N on the continuum at $8500 \AA$ of
  	$S/N=5$ for $i_{AB}=25$, and $S/N=5$ for an emission line with a flux $f_{\lambda}= 1.5\times 10^{-18} erg/s/cm^2/\AA$.
               
               With these criteria we selected 246 galaxies. 
               From this sample  we excluded 37 galaxies due to spectral defects or strong residuals from skyline subtraction in the LyC region of the spectrum ($870-910\,\AA$ rest-frame), and 1 galaxy with possible AGN features. We also discarded 6 further objects whose spectra were clearly contaminated by very bright neighbours, after visually checking the 2-dimensional spectra and the available low resolution images.  Finally, given that even a few spurious excess in the LyC region of the spectrum can affect our measurements, we checked that the 2-dimensional spectrum of each source was not contaminated by higher-order spectra coming from other slits in the same mask. We found only one source that could present this kind of contamination and we excluded it from the sample.  
               
               The final sample contains  201 galaxies. 106 of these are in the COSMOS field, 22 in the ECDFS field and 73 in the VVDS-2h field. The redshift and R magnitude distributions of the galaxies in the final sample,  with the scatter plot between these two quantities, are shown in Fig. \ref{fig:histoz}. The median redshift and R magnitude are $z=3.81$ and $R=24.91$, respectively.

               In \cite{marchi} and \cite{lucia} a careful cleaning procedure based on multi-band high resolution HST imaging was applied, to exclude all the galaxies contaminated by any possible interloper. In these papers we found contamination fractions of $28\%$ and $52\%$, respectively. This procedure cannot be applied  on the  present sample since the availability of multi-color HST imaging  would restrict the application of this procedure only to a very small subset ($\sim 35$ galaxies) reducing the sample significantly. Instead, we simply excluded the clearly contaminated objects as previously explained in this section. We assume, for the rest of the analysis   that all the sub-samples that we will define in the next section, contain a statistically similar contamination.  
We discuss the effects of the contamination from faint low-redshift interlopers  on our samples in Section \ref{sec:contamination} and in the Appendix.
         \section{Method}
         \label{sec:method}
         \subsection{Ly$\alpha$ and UV properties}
         \label{sec:properties}
\begin{figure*}
				\centering
					\includegraphics[width=0.45\linewidth]{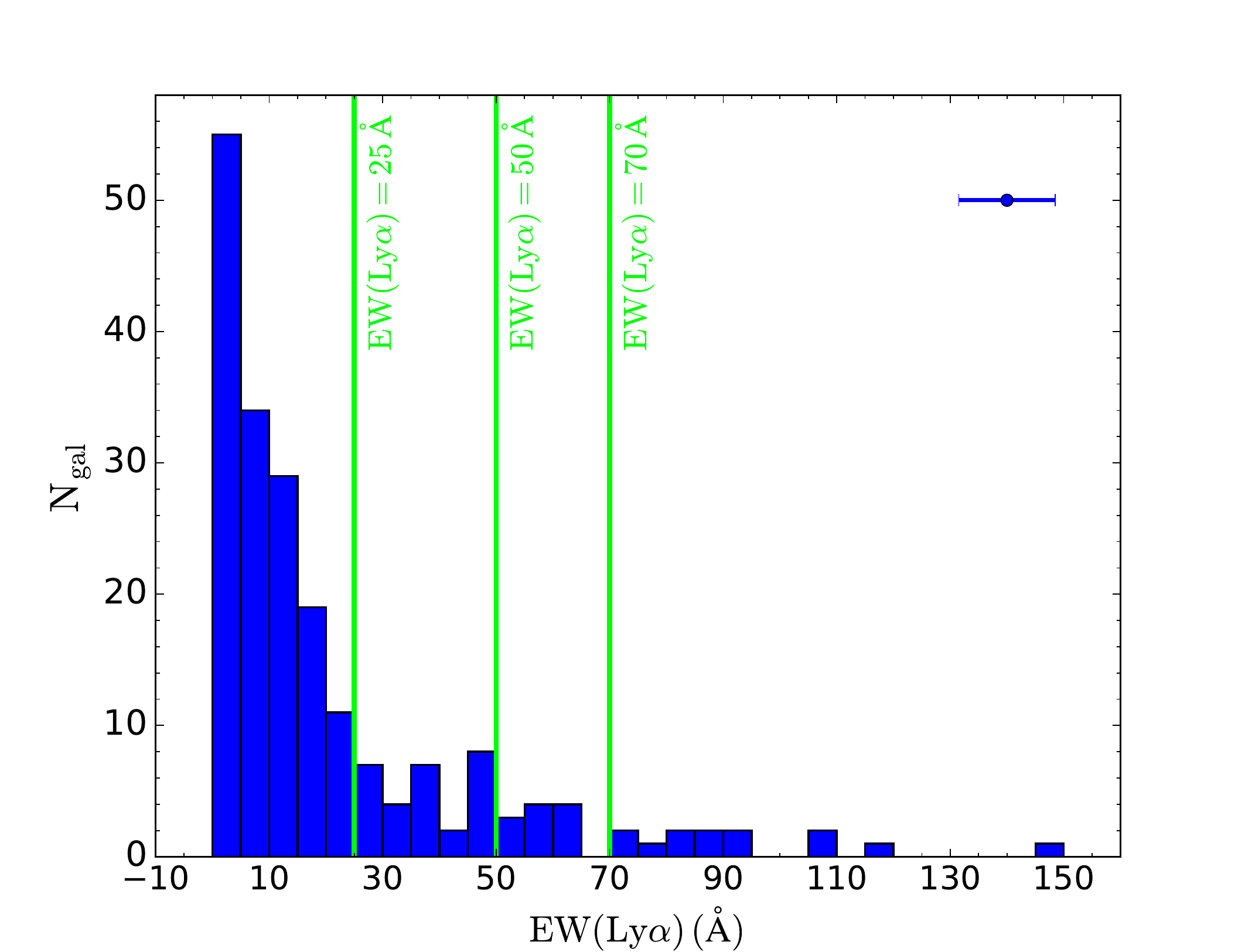}
					\includegraphics[width=0.45\linewidth]{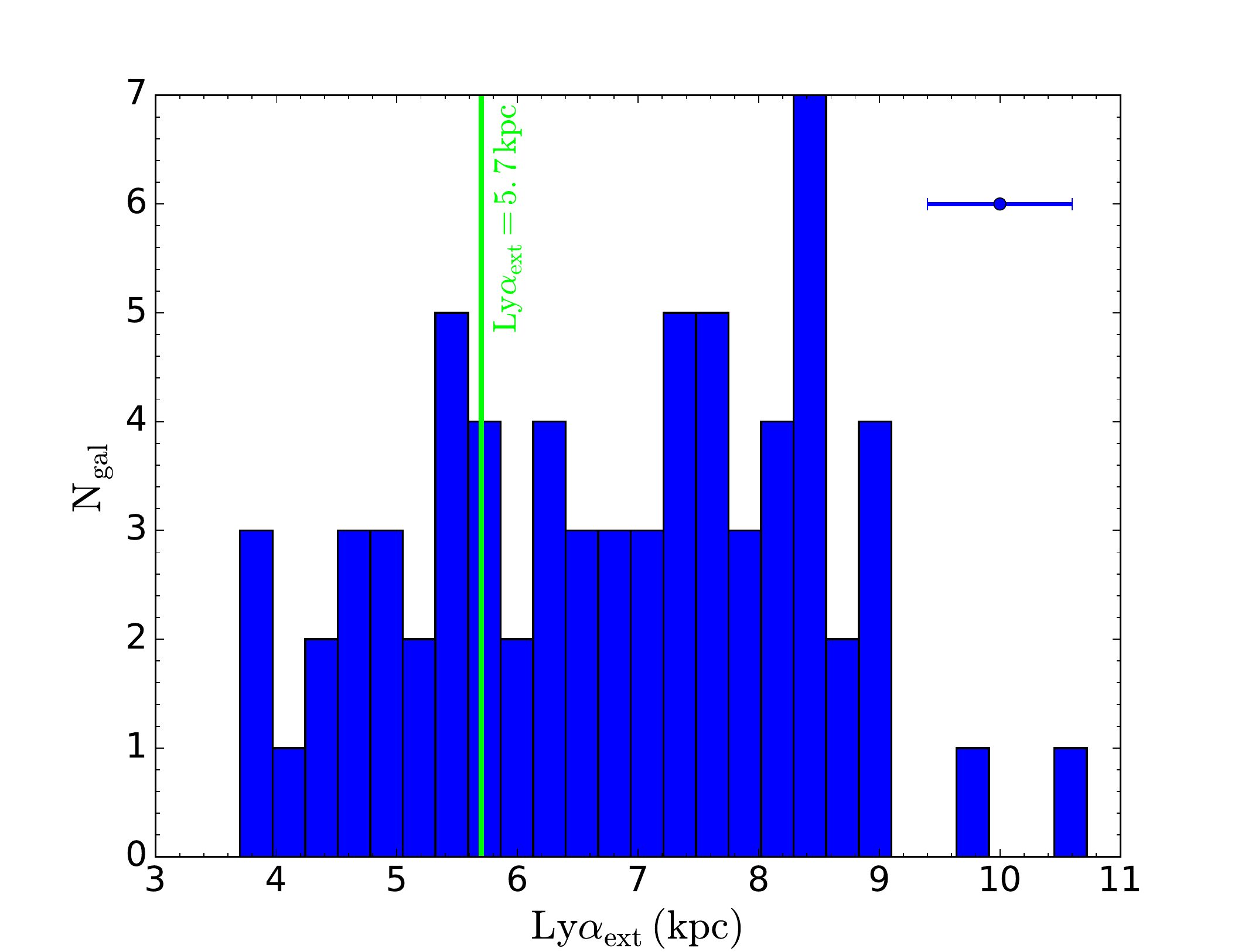}\\
					\includegraphics[width=0.45\linewidth]{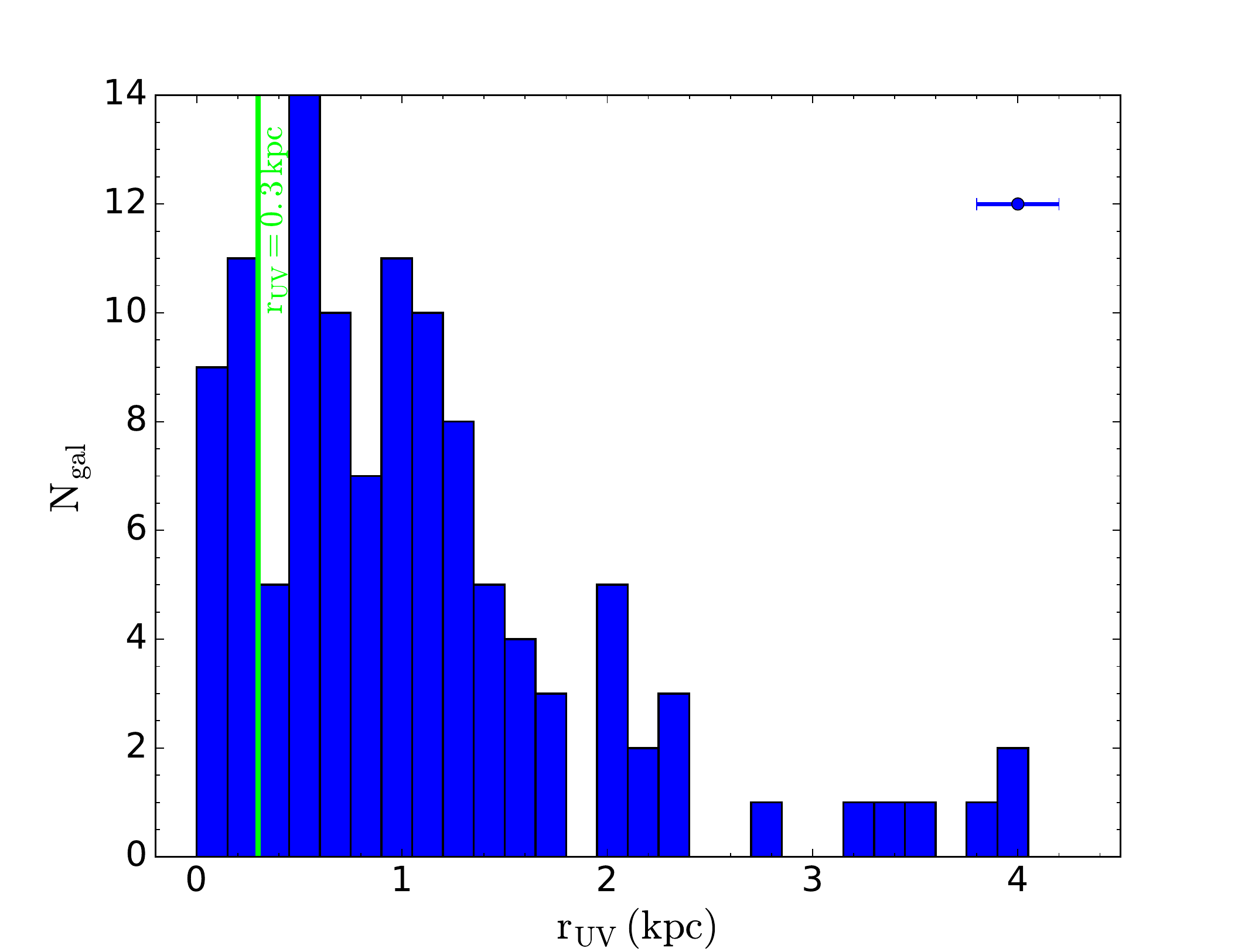}
					\includegraphics[width=0.45\linewidth]{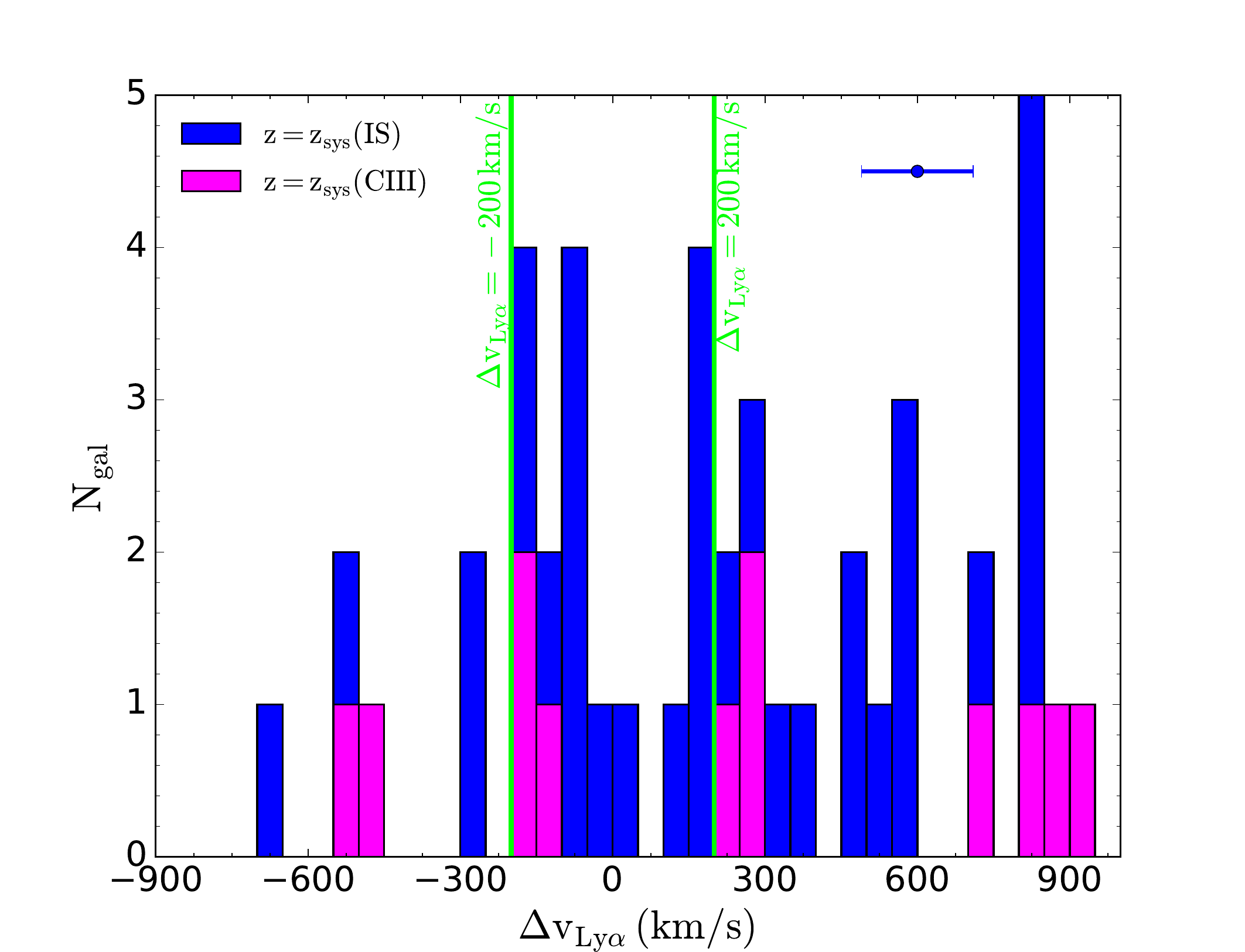}
					\caption{Distributions of the parameters that we analysed in this work (see Section \ref{sec:properties}). The green vertical lines correspond to the cuts that we applied to select the sub-samples (see Section \ref{sec:sub}). Note that we left out one source with $EW(Ly\alpha)=269\,\AA$ from the EW(Ly$\alpha$) histogram for clarity. We also color coded the distribution of the Ly$\alpha$ velocity offset according to the line used to measure $z_{sys}$: magenta indicates $z_{sys}$ from the CIII(1907.07$\AA$) line and  blue indicates the $z_{sys}$ from the interstellar absorption lines. In each panel we show the average error on the relevant parameter in the upper right. Note that for the EW(Ly$\alpha$) we show the error for $EW(Ly\alpha)=50\, \AA$; for the Ly$\alpha_{ext}$ we plot for reference an error of $0.6\,kpc$, which corresponds to a $10\%$ error of the value used to divide the samples.}
					\label{fig:EW}
			   \end{figure*}

        The indirect indicators of LyC emission that we can in principle use exploiting the VUDS data, are: the  Ly$\alpha$ EW and the Ly$\alpha$ FWHM \citep{dijkstra16}, the Ly$\alpha$ velocity offset with respect to the systemic redshift \citep{dijkstra16,verhamme15}, the Ly$\alpha$ extension (Verhamme et al. in prep) and the UV compactness \citep{wise09,izotov16}. In this section we describe the methods  used to evaluate these quantities and their errors for the galaxies in our sample.\\
         \begin{itemize}
         	\item \emph{Ly$\alpha$ EW}
         	
         	We use the values of the Ly$\alpha$ EW from \cite{cassata:EW}, that are evaluated using a continuum and line flux  estimate from the IRAF splot tool.
         	The EW(Ly$\alpha$) distribution is shown in Fig. \ref{fig:EW}. $\sim10\%$ of the sample has $EW(Ly\alpha)\geq55\, \AA$ and $\sim25\%$ has $EW(Ly\alpha)\geq25\, \AA$ in agreement with the statistics at this redshift \citep[e.g.][]{shapley03,stark10}. We note  that these fractions are slightly higher than those found by \cite{cassata:EW}, also with the VUDS data. This is due to the fact that we included in our sample only the spectra with quality flag 3 and 4 \citep[probability greater than 95\% for the spectroscopic redshift to be correct, see][]{lefevreVIMOS} while \cite{cassata:EW} used also galaxies with a less secure spectroscopic redshift which usually do not show  Ly$\alpha$ in emission. As already explained in the previous section however, we decided to include only the galaxies with secure spectroscopic redshifts in our sample to avoid the presence of spurious objects in our redshift range. 
         	
         	The errors on the EW(Ly$\alpha$) were evaluated following Eq. 7 in \cite{vollmann06} and are on average $\sim 17\%$ of the measured values.
         	

         	\item \emph{Ly$\alpha$ velocity offset} ($\Delta v_{Ly\alpha}$)
         	
         	To estimate the Ly$\alpha$ velocity offset with respect to the systemic redshift, a good knowledge of the systemic redshift is needed: this  can be obtained  either  from photospheric stellar absorption lines (such as OIV($1343\,\AA$) and SiIII($1417\,\AA$)) which are too weak in our individual spectra,  or from  nebular emission lines. Only 12 galaxies in our  sample show CIII(1907.07$\AA$) in emission: in these cases the systemic redshift  was estimated from  the centroid of the line. We note that, since the VUDS resolution does not allow to distinguish the CIII(1907.07$\AA$) doublet, we used a single Gaussian fit to estimate the centroid. For 33 further sources, that are characterized by strong SiII(1260.42$\AA$), CII(1334.53$\AA$) and/or SiII(1526.71$\AA$) in absorption, we applied the relation found by \cite{steidel10} between the  redshift of the inter-stellar absorption lines ($z_{IS}$) and the systemic redshift: 
         	\begin{equation}
         	z_{sys}=z_{IS}+0.00299-0.00291(2.7-z_{IS})
         	\label{eq:steidel}
         	\end{equation}
         	that was derived analysing a sample of 86 galaxies at $z\sim2.3$ for which the systemic redshift was known from the H$\alpha$ emission line. 
         	We assume here that this relation is also valid at the redshifts of our sample (though see the discussion in the next paragraph). 
         	We then measured the center of the Ly$\alpha$ line and  estimated   $\Delta v_{Ly\alpha}$ with respect to the systemic. 
         	
         	The distribution of the Ly$\alpha$ velocity offset is shown in Figure \ref{fig:EW}. It presents an extended tail to negative velocities, which are not commonly observed in star-forming galaxies, both at low and high redshift \citep[e.g.][]{guaita13,mclinden14,song14}. We could not test the \cite{steidel10} relation at our redshifts, so it might be that these negative velocities come from a bad evaluation of the systemic redshifts from the IS absorption lines. However, as shown in Figure \ref{fig:EW}, this is likely not the case, since there are galaxies with negative velocity where the systemic redshift was evaluated from the CIII(1907.07$\AA$) emission. We also note that \cite{gronke17} recently found that a small fraction of Ly$\alpha$ emitters from the ’MUSE-Wide’ survey shows negative velocities \footnote{Note that \cite{gronke17} defines the velocity offset as $\Delta v_{Ly\alpha}=c\frac{z-z_{Ly\alpha}}{z+z_{Ly\alpha}}$, so their positive offsets correspond to negative values according to our definition.} up to $\sim -300\,km/s$, although their systemic redshifts were derived in a
         		different way, namely from a \emph{shell model} fit of the Ly$\alpha$ line.

         		The errors on the velocity shifts were  evaluated summing quadratically the uncertainties on the determination of the Ly$\alpha$ centroid and the centroid of the CIII(1907.07$\AA$) emission or IS absorption lines. The individual centroid errors depend on the S/N at the peak of the line, the resolution of the spectrum ($\sim 1300 km/s$) and on a constant that depends on the kind of noise, that we assume $1/1.46$ as in \cite{lenz92}. For example, for an average $S/N\sim5$ both at the peak of the Ly$\alpha$ and at the peak of the CIII(1907.07$\AA$) or at the bottom of the IS absorption lines, we obtained individual errors of $\sim 80km/s$, and a final error of $\sim 110km/s$. We note that for very bright lines, the S/N is higher than 5 and therefore the uncertainty on the Ly$\alpha$ velocity shift is slightly lower.

         	\item \emph{Ly$\alpha$ spatial extent} (Ly$\alpha_{ext}$) 
         	
         	The Ly$\alpha$ spatial extent was evaluated directly from the 2-dimensional spectrum of the sources. We collapsed the 2-d spectrum in the wavelength range of the Ly$\alpha$ and then applied a Gaussian fit along the y-axis perpendicular to the spectral dispersion. The  Ly$\alpha$ spatial extent was then evaluated as the FWHM of the best fit. We could measure  this parameter for the 70 galaxies with $EW(Ly\alpha)>10\,\AA$ that did not present any skylines close to the position of the Ly$\alpha$ in the spectrum. For the objects with lower EW(Ly$\alpha$) the S/N was not sufficient to derive the measurement. We note that our estimates of the Ly$\alpha_{ext}$ cannot be taken as absolute estimates of this quantity since we did not deconvolve it by  the Point Spread Function (PSF) of the
         	observations. However, this is not an issue for our analysis since we are  only interested  in the relative  Ly$\alpha$ extensions and, since the observations were taken under very similar conditions, we can assume that the PSF affects all the sources in the same way.
         	The distribution of the Ly$\alpha$ spatial extent is shown in Fig. \ref{fig:EW}.
         	
         	Given that the $Ly\alpha_{ext}$ is measured only on objects with bright Ly$\alpha$, the main uncertainty involved in its estimate   is
         		the choice of the background in the 2-dimensional
         		spectrum, before fitting the Gaussian profile. The reason is that this background is estimated on relatively few pixels, due to the limited length of the slits in the spatial direction. We therefore allowed the background
         		to vary within the 2$\sigma$ error of its mean value
         		for 1000 times, and then fitted again a Gaussian profile to the line. The errors are finally obtained from the dispersion of the FWHM distribution and vary from $0.1\, kpc$ to a maximum of $1\, kpc$, with an average value of $\sim 10\%$ of the measured $Ly\alpha_{ext}$.
         	
         	\item \emph{UV rest-frame morphologies (r$_{UV}$)}
         	
		     We used the effective radii obtained by \cite{ribeiro16} with GALFIT for the rest-frame UV continuum sizes. These measurements were available  for the 115 objects covered by deep I band imaging (HST, F814W) in the COSMOS and ECDFS fields (U-band rest-frame). The distribution of these values is show in Fig. \ref{fig:EW}. We note that we took only the objects for which GALFIT converged. The errors on r$_{UV}$ were also evaluated with GALFIT and are on average $\sim0.2\, kpc$.
      
         \end{itemize}	   
               In principle also the FWHM of the Ly$\alpha$ line would be measurable from the spectra. However, models predict that the LyC emitters are those with very narrow emission \citep[$\sim 200\, km/s$ according to][]{dijkstra16}, which are not measurable in the low resolution VUDS spectra ($R\sim 230$).     
			   
\subsection{The escape fraction of LyC photons}
 \label{sec:fesc}
The escape fraction of LyC photons is the fraction of ionizing radiation that is able to escape from the galaxy into the IGM without being absorbed, relative to the total number of photons produced \citep{fesc,fesc2}, and references therein). Determining this quantity, known as 
the \emph{absolute escape fraction}, $f_{esc}$, requires knowledge of the intrinsic number of
ionising photons produced by the galaxy itself. However, the intrinsic spectral energy distribution (SED) of a galaxy is not known a priori, especially in the rest-frame far UV where dust reddening could be severe. 
               
A related quantity, more used in observational studies, is the \emph{relative escape fraction},  which is defined as \citep{steidel,siana}:

\begin{equation}
\label{eq:fesc}
f_{esc}^{rel}(LyC)=\frac{L_{\nu}(1470)/L_{\nu}(895)\cdot f_{\nu}(895)/f_{\nu}(1470)}{e^{-\tau_{IGM,z}}}
,\end{equation}
where $L_{\nu}(1470)/L_{\nu}(895)$ is the ratio of the intrinsic luminosities
at 1470 and 895 $\AA$ rest frame and $f_{\nu}(895)/f_{\nu}(1470)$ is the ratio of the
observed flux densities at the same wavelengths. $e^{-\tau_{IGM,z}}$ is the \emph{transmissivity} that takes into account the photoelectric absorption of photons with $\lambda \leq 912 \AA$ by the IGM and that then depends on the redshift.  

The intrinsic luminosity ratio, $L_{\nu}(1470)/L_{\nu}(895)$, depends on the
physical properties of the galaxies, such as the mean stellar ages,
metallicities, stellar initial mass functions (IMFs), and star formation histories (SFHs).  Under reasonable assumptions, it is usually taken as 3-5 \citep[see][and the discussion in Section \ref{results}]{lucia}. The transmissivity can be estimated by simulating several absorbers in different lines of sight and evaluating the mean intergalactic attenuation curve \citep{inoue:transmissivity,worseck14}. We highlight here that there is  a large scatter in the IGM transmission around the mean at each given redshift, as shown, for example, in Fig. 2 of \cite{vanzella15} and as computed directly from spectral fitting \citep{thomas17}.

 The quantity that we can measure directly  from the VUDS spectra is the flux density ratio ($\frac{f_{\nu}(895)}{f_{\nu}(1470)}$) that we use in this paper as an indicator of LyC emission. We explore the caveats of this assumption in Section \ref{results}.

Since the single spectra have low S/N at LyC wavelengths and  cannot give precise values   of the LyC signal, to increase the sensitivity of our measures, we grouped the galaxies in different sub-samples according to the properties derived in the previous section, and then, for each sub-sample, we produced   stacked spectra where we measured directly  the ratio between the LyC and the UV continuum density fluxes, $\frac{f_{\lambda}(895)}{f_{\lambda}(1470)}$,   as explained in Section \ref{sec:stack}. We note that in our previous paper \citep{marchi}, we tried to evaluate the flux density ratio from the individual spectra and obtained a tentative trend between the flux density ratio and the Ly$\alpha$ EW for the galaxies in our sample. However, those values were only upper limits and had large errors. For this reason, we do not attempt here to do the same procedure, but rely on the stacks of sub-samples selected according to the observed properties to have better estimates of this quantity.
\subsection{Definition of the sub-samples}
\label{sec:sub}
 In this Section we explain the criteria that we used to select the different sub-samples. Given that the properties we want to explore were measured in different subsets (e.g. $\Delta v_{Ly\alpha}$ in 45 galaxies, r$_{UV}$ in 115 objects and so on) also the division in sub-samples according to the galaxies' properties is different. 
 
\begin{itemize}
	\item \emph{EW(Ly$\alpha$) sub-samples}
	
	Theoretical predictions and observational studies in the local universe \citep[e.g.][]{dijkstra16,verhamme17} suggest that the escape fractions of Ly$\alpha$ and LyC photons are correlated. Since high Ly$\alpha$ escape fractions generally imply high Ly$\alpha$ EWs, we can use this quantity to test the correlation between Ly$\alpha$ and LyC radiation. We measured the  EW(Ly$\alpha$) for all the 201 galaxies in the final sample. For this reason, we could study the dependence of the flux density ratio with the EW(Ly$\alpha$) imposing different cuts on this quantity and see how the LyC emission changes increasing the EW(Ly$\alpha$) cut. We start with a cut at $EW(Ly\alpha)=25\,\AA$, to include in the sub-sample all the LAEs, and then we increase it to $EW(Ly\alpha)=50\,\AA$ and finally to $EW(Ly\alpha)=70\,\AA$. We therefore selected and stacked all the galaxies with $EW(Ly\alpha)\ge25\,\AA$, $EW(Ly\alpha)\ge50\,\AA$ and $EW(Ly\alpha)\ge 70\,\AA$ and finally compared the values of $\frac{f_{\lambda}(895)}{f_{\lambda}(1470)}$ obtained for these samples with those of their complementaries.
	
		 \item \emph{$\Delta v_{Ly\alpha}$ sub-samples}
		 
		 According to theoretical predictions \citep{dijkstra16,verhamme15}, the galaxies with a higher probability to leak LyC emission, have a   Ly$\alpha$ emission line that emerges  very close to the systemic redshift \citep[for example $\Delta v_{Ly\alpha}<150\,km/s$ according to][]{verhamme15}. 
We therefore  selected and stacked all the sources with Ly$\alpha$ velocity offset around zero, $-200\leq\Delta  v_{Ly\alpha}\leq 200\,km/s$, and those with $\Delta v_{Ly\alpha}>200\,km/s$ and $\Delta v_{Ly\alpha}<-200\,km/s$. We had to select galaxies in a larger range of $\Delta  v_{Ly\alpha}$ with respect to the predicted values because  we would have had too few galaxies in the interval $|\Delta v_{Ly\alpha}|<150\,km/s$ to do a reliable stack.
		 
	\item \emph{Ly$\alpha_{ext}$ sub-samples}
	
	We did not find in the literature any study focused on the relation between the Ly$\alpha$ emission size and the presence of LyC radiation, that could drive our sample division. The distribution of the observed sizes is also more or less flat between 4 and 9 kpc (see Figure \ref{fig:EW}). Recently \cite{yang16} found an anti-correlation between the Ly$\alpha$ extension and the  Ly$\alpha$ escape fraction. Since we expect a correlation between Ly$\alpha$ and LyC escape fractions \citep{dijkstra16}, we suppose that the Ly$\alpha$  compactness could favour the escape of LyC radiation.  To investigate this scenario, we selected the 20 most compact sources in Ly$\alpha$ to see if this sample has a higher LyC emission compared to the galaxies with a larger Ly$\alpha$ extension. We selected a sub-sample with $Ly\alpha_{ext}\leq 5.7\,kpc$ and a complementary sample with $Ly\alpha_{ext}> 5.7\,kpc$. We chose 20 objects because it is the minimum number required to produce a stack not dominated by the errors.
	
	\item \emph{$r_{UV}$ sub-samples}
	
		\cite{izotov16} demonstrated that a selection for UV compact objects, along with a selection for high  $[OIII]\lambda  5007/[OII]\lambda3727$ emitters, appears to pick up very efficiently LCEs in the local universe. We want to test this correlation at higher redshifts. The only indication that we have at $z\sim 3$ is the size of the well known LCE \emph{Ion2} \citep{vanzella16,debarros17}, with $r_{UV}= 0.3\, kpc$. Since we have a sufficient number of objects with $r_{UV}\le 0.3\, kpc$ (20 sources), we decided to define this as a sub-sample and, to investigate the dependence between the UV radius and the LyC signal, compare its flux density ratio with that of the complementary sample ($r_{UV}> 0.3\, kpc$).
\end{itemize}

	\begin{table*}
		\centering
		\begin{tabular}{lc cc}
			\toprule
			\toprule
			
			Sub-sample & Number of sources & Complementary sample & Number of sources \\
			
			\midrule
			$EW(Ly\alpha)\ge25\,\AA$ & 52 & $EW(Ly\alpha)<25\,\AA$ & 149 \\
			$EW(Ly\alpha)\ge50\,\AA$ & 25 & $EW(Ly\alpha)<50\,\AA$ & 176\\
			$EW(Ly\alpha)\ge70\,\AA$ & 14 & $EW(Ly\alpha)<70\,\AA$ & 187\\
			$Ly\alpha_{ext}\le 5.7\,\mathrm{kpc}$ & 20 & $Ly\alpha_{ext}>5.7\,\mathrm{kpc}$ & 50 \\
			$|\Delta  v_{Ly\alpha}|\le 200\,\mathrm{km/s}$ & 17 & $|\Delta v_{Ly\alpha}|>200\,\mathrm{km/s}$ & 28 \\
			$r_{UV}\le0.3\,\mathrm{kpc}$ & 20 & $r>0.3\,\mathrm{kpc}$ & 95\\
			
			\bottomrule[1.5pt]
		\end{tabular} 
		\vspace{0.1cm}
		\caption{Number of sources in each sub-sample considered. (1) Selection criterion for the sub-sample; (2) Number of sources in the sub-sample; (3) Selection criterion for the complementary sub-sample; (4) Number of sources in the complementary sub-sample.}
		\label{tab:samples}
	\end{table*}

	We summarise in Table \ref{tab:samples} the division in different sub-samples and complementary samples, showing also the number of galaxies in each sample. In Figure \ref{fig:EW}, the green vertical lines correspond to the values that we chose to select the different sub-samples.

\subsection{Stacking procedure and $f_{\lambda}(895)/f_{\lambda}(1470)$ evaluation}
\label{sec:stack}
 We followed the same stacking procedure described in \cite{marchi}.  We first shifted each
    one-dimensional spectrum to its rest-frame and normalised it using its mean value in the wavelength range $1420-1520 \, \AA$ where no particular features
     are present. To bring the spectra rest-frame, we used the systemic redshifts, for the sources that presented inter-stellar absorption lines or the CIII(1907.07$\AA$)  line (45 galaxies, see Sec \ref{sec:properties}), and the VUDS official redshifts for the other sources \citep[see][for details on the redshift evaluation]{lefevreVIMOS}. To take into account the noise of each spectrum, we computed the stack as a weighted average of the normalised spectra in each sub-sample, using the statistical errors of the individual flux density ratios $\frac{f_{\lambda}(895)}{f_{\lambda}(1470)}$ 
     as weights \footnote{The errors have been evaluated with the classic errors propagation from the standard deviations around the mean values in the LyC and UV ranges ($870-910\AA$ and $1420-1520\AA$, respectively).}. To make the average, we resampled each spectrum to the same grid that goes from $870\,\AA$ to $1700\,\AA$ with a step of $5.355/(1+z_{median})$. $5.355$ is the nominal VIMOS Low Resolution $\AA/pixel\, scale$ of the observed-frame spectra, and $z_{median}$ is the median redshift of the sub-sample.
     
     The flux density ratio  $f_{\lambda}(895)/f_{\lambda}(1470)$  can be measured directly from the stack, averaging the signal in the interval $870-910\, \AA$. However, given the very small size of some of our sub-samples, taking this value simply from the stack would not give a reliable estimate of the flux density ratio and its error. For this reason, we used the bootstrapping technique to
     estimate the  uncertainty on this quantity. Using this approach
     does not require making any assumptions on the distribution of our data. 
     For each sub-sample of N galaxies, we therefore created 5000 realizations of it, randomly extracting N galaxies with replacement. We then stacked each realization and computed the flux density ratio. We finally evaluated the mean and the $68\%$ confidence level of each $f_{\lambda}(895)/f_{\lambda}(1470)$ distribution. 

	\section{Results}
	\label{results}
\begin{figure*}
		\centering
		\includegraphics[width=1\linewidth]{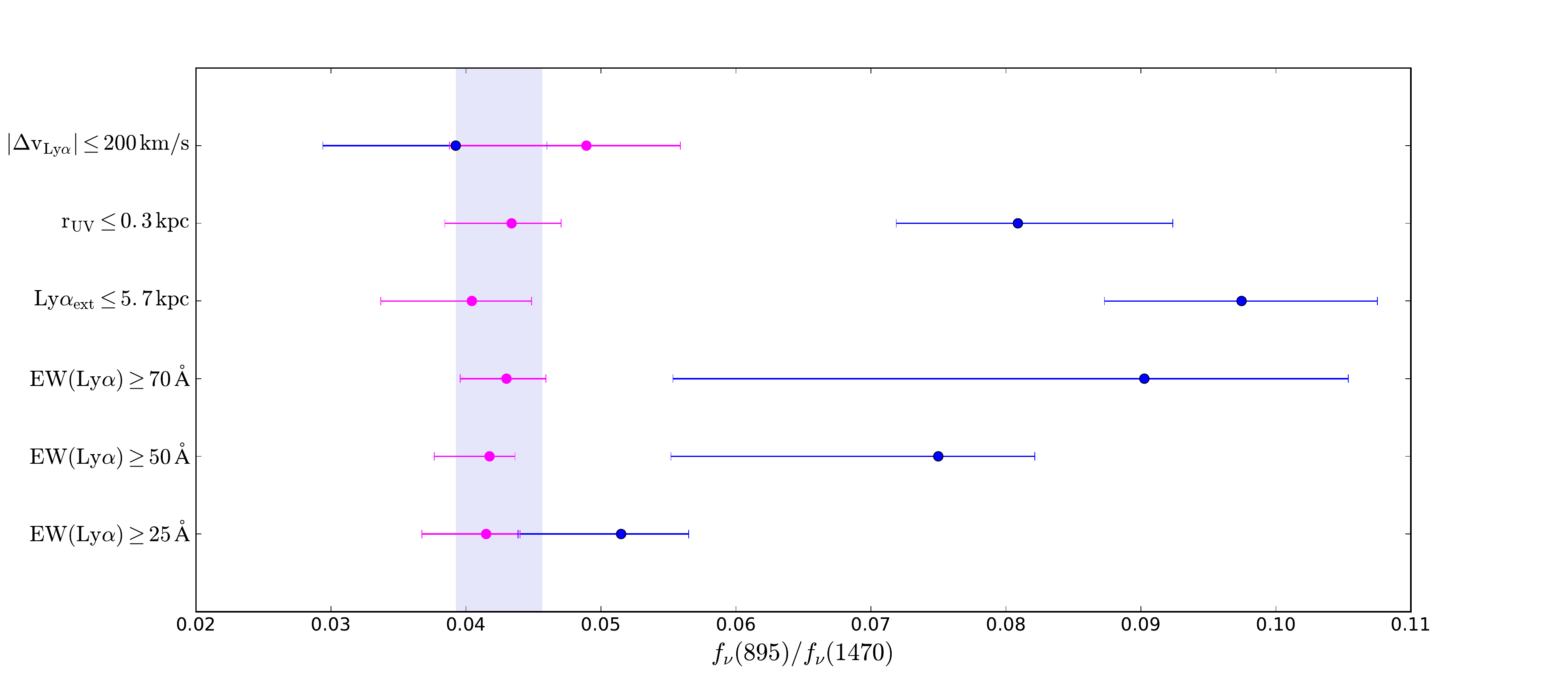}
		\caption{Flux density ratios evaluated from the stacks of the samples in the y-axis (blue dots) and from the complementary samples (magenta dots) as indicated in Table \ref{tab:samples}. The lavender vertical band is the 1$\sigma$ confidence interval evaluated for the total sample of 201 galaxies.}
		\label{fig:fratio}
	\end{figure*}

 The flux density ratios representative of each sub-sample, measured as explained
	in Section \ref{sec:stack} and converted from $f_{\lambda}$ to $f_{\nu}$, are shown in Figure \ref{fig:fratio}. The blue dots are the values corresponding to the sub-samples defined in the y-axis and the magenta dots are those corresponding to the complementary samples as indicated in Table \ref{tab:samples}. The lavender vertical band is the 1$\sigma$ confidence interval for the total sample of 201 galaxies. We expect that the average signal of LyC is almost washed away in the total sample, given that this is probably dominated by sources with no LyC leakage, and we assume, for the moment, that the signal $\frac{f_{\nu}(895)}{f_{\nu}(1470)}$  observed in the total sample basically comes from contamination from low redshift interlopers. 
	This will be further discussed in Section \ref{sec:contamination}.
	
	We can see from Figure \ref{fig:fratio} that most of the sub-samples (blue dots) show a $\frac{f_{\nu}(895)}{f_{\nu}(1470)}$  much larger than the $1\sigma$ interval obtained for the entire sample (lavender band) and that their complementary samples (magenta  dots) are instead in agreement with it. The only exceptions are the $\Delta v_{Ly\alpha}$ samples that we better discuss later.
	

The parameters which appear to (anti)correlate the strongest with the flux density ratio are EW(Ly$\alpha$), Ly$\alpha_{ext}$ and $r_{UV}$. 
The higher is the Ly$\alpha$ equivalent width, the higher is the flux density ratio, and the smaller are the Ly$\alpha$ spatial extent and the UV effective radius, the higher is the LyC radiation. In particular, the sub-samples with $Ly\alpha_{ext}\le5.7\,kpc$ and  $r_{UV}\le0.30\,kpc$ have values for the flux density ratio that are more than $2\sigma$ higher than their complementaries. 

 We note that the uncertainties in the flux density ratio of some sub-samples are particularly large. This is because some of these sub-samples contain very few sources (in particular the sub-sample with $EW(Ly\alpha)\geq70\,\AA$ contains only 14 galaxies). 
        
\begin{figure*}
		\centering
		\includegraphics[width=0.75\linewidth]{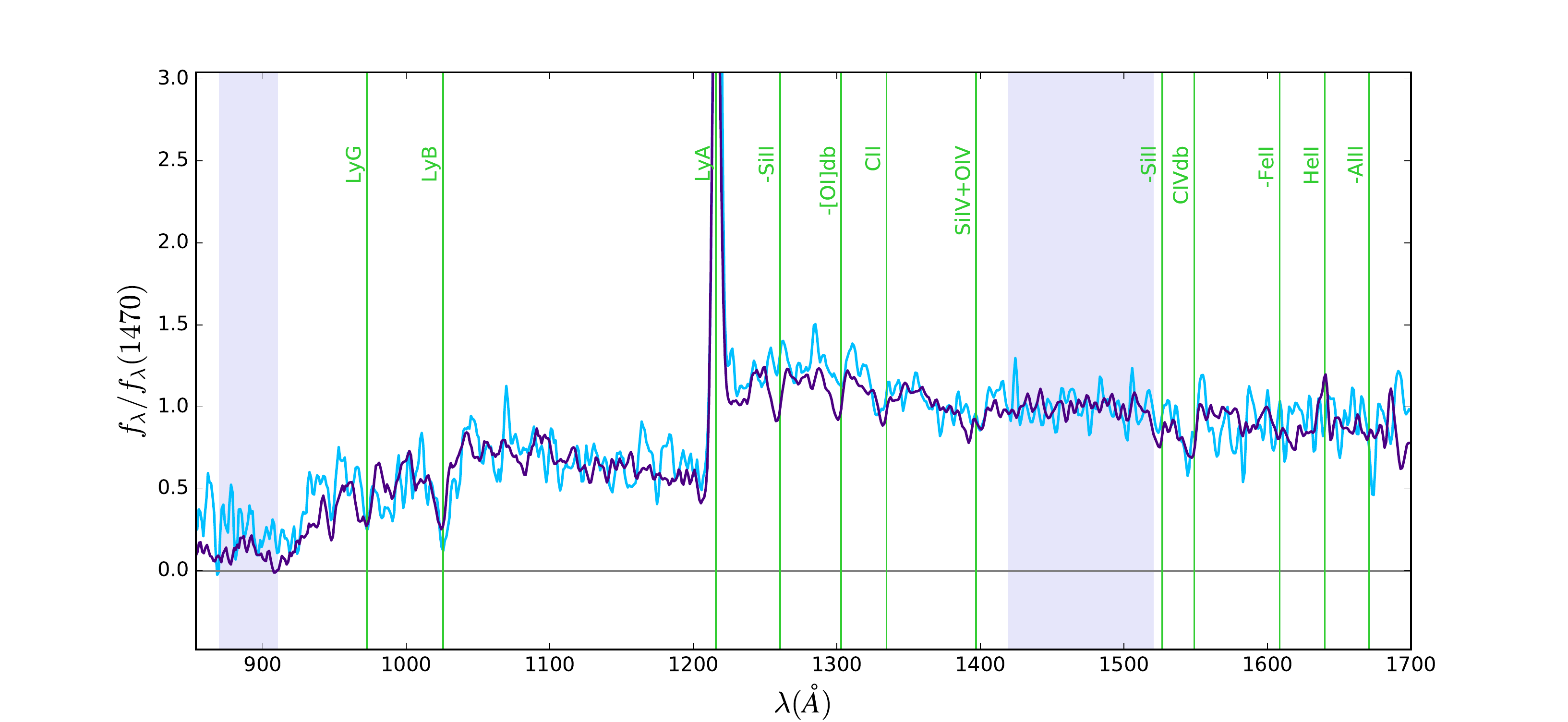}
		\caption{Comparison of the spectral stacks of the sub-samples with $Ly\alpha_{ext}\le5.7\,kpc$ (cyan line) and with $Ly\alpha_{ext}>5.7\,kpc$ (purple line). The vertical lavender bands indicate the LyC range ($870-910\,\AA$) and the UV range ($1420-1520\,\AA$) where we have normalised each spectrum. The signal in the LyC range in the stack of the sub-sample with $Ly\alpha_{ext}\le5.7\,kpc$ is about 2.5 times higher than that of the sub-sample with $Ly\alpha_{ext}>5.7\,kpc$. The spectra in the Figure have been smoothed by 3 times the step of the stacked spectrum, which is $1.06\,\AA$ for the stack of the sub-sample with $Ly\alpha_{ext}\le5.7\,kpc$ and $1.12\,\AA$ for the sub-sample with $Ly\alpha_{ext}>5.7\,kpc$, to emphasize the difference in the LyC region between the two sub-samples.}
		\label{fig:stackLyAext}
	\end{figure*}

        In Figure \ref{fig:stackLyAext} we show the normalised spectral stacks of the sub-samples with $Ly\alpha_{ext}\le5.7\,kpc$ (cyan line) and with $Ly\alpha_{ext}>5.7\,kpc$ (purple line) for reference. 
	The excess in the LyC range is clearly visible even from the simple stack in the sub-sample with small Ly$\alpha_{ext}$, while a much lower  signal is present in the stack of the sub-sample with higher Ly$\alpha_{ext}$ in this range.  We emphasize  that we are looking for a differential LyC signal between the sub-samples, because at this stage, we are not taking into account the contamination from lower-redshit interlopers and we are assuming the same statistical contamination in all the sub-samples. 
	
\begin{figure}
\centering
\includegraphics[width=0.9\linewidth]{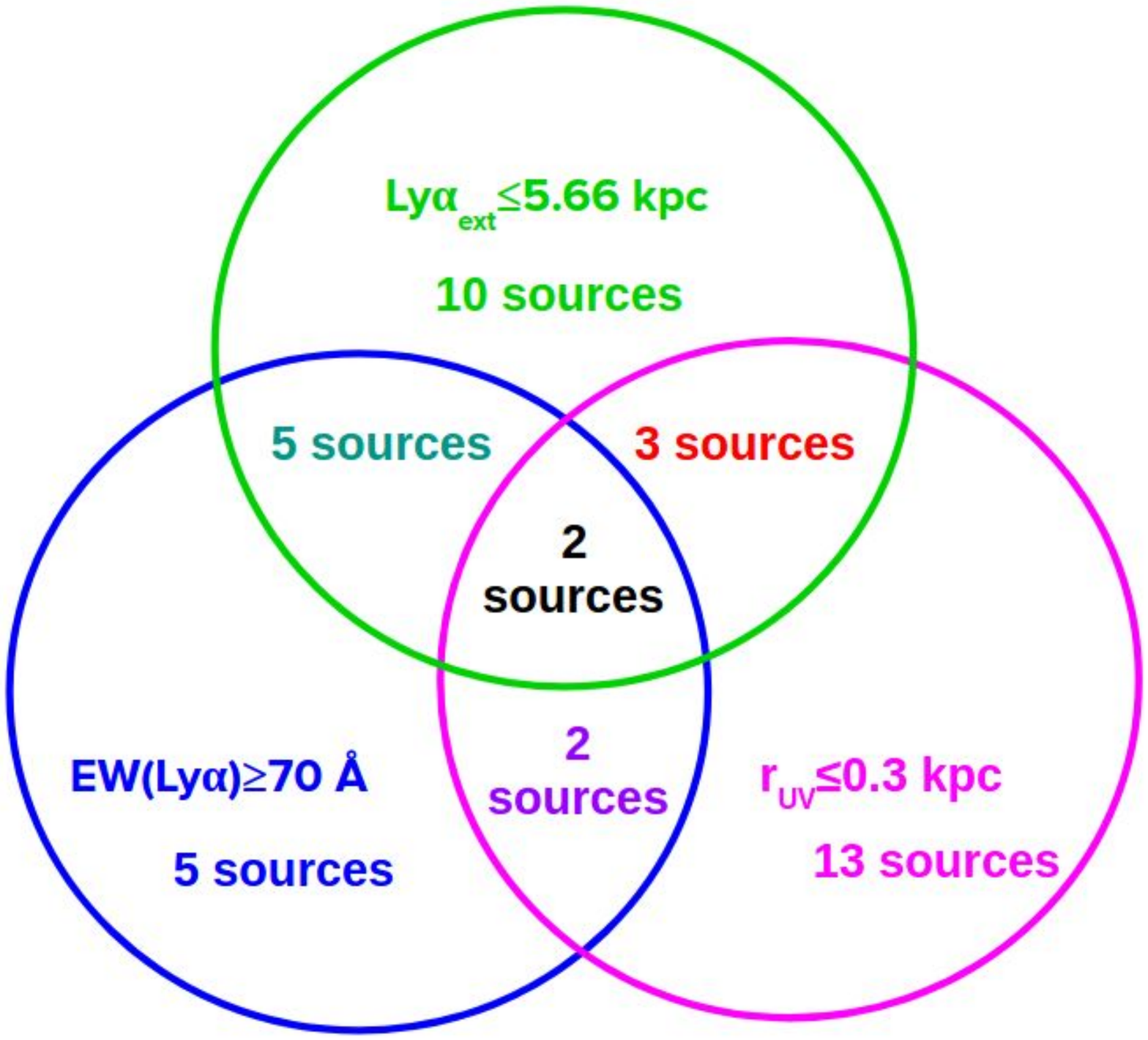}
\caption{Venn diagram showing the number of sources contained in each significant sub-sample ($EW(Ly\alpha)\ge 70\,\AA$, $Ly\alpha_{ext}\le 5.7\,kpc$ and $r_{UV}\le0.30\,kpc$) and the sources in common.}
\label{fig:common}
\end{figure}

The spectral stacks for the three most significant parameters ($EW(Ly\alpha)\ge 70\,\AA$, $Ly\alpha_{ext}\le 5.7\,kpc$ and $r_{UV}\le0.30\,kpc$) are not independent, since many objects belong to more than one of these sub-samples.
We show in Figure \ref{fig:common} the Venn diagram with the number of sources contained in each sub-sample and the sources in common. 
We could not measure all the parameters for all the galaxies in the final sample, so the sources in common between the three sub-samples could be more than the present numbers.
We are not able to study the LyC properties of the sources in common between the sub-samples because the number of galaxies that satisfy all the three conditions is too low (see Figure \ref{fig:common}) and the stack is dominated by noise.

To test the correlation between the different parameters, we evaluated the Spearman's rank correlation coefficients between the sub-samples for which we measured EW(Ly$\alpha$), Ly$\alpha_{ext}$ and r$_{UV}$.
We found that the most correlated parameters are, as expected,  the Ly$\alpha$ spatial extension and the UV effective radius, with a coefficient of 0.32, that implies that the smaller is r$_{UV}$, the smaller is $Ly\alpha_{ext}$. 
We also found a  relatively weak  anti-correlation between the Ly$\alpha$ spatial extension and the EW(Ly$\alpha$), in the sense that the galaxies with lower EW(Ly$\alpha$), present a more extended Ly$\alpha$ profile. In principle this weak trend could be due to an  improper extraction of
the 1-dimensional spectra for objects with large Ly$\alpha$ spatial extension
because   the extraction aperture is determined from the UV portion of the
spectra. Therefore, for extended Ly$\alpha$ emission, part of the Ly$\alpha$  flux could be
lost and lower EW could be  measured from the 1-dimensional spectrum. However, we 
checked that this is not the case for our galaxies by re-extracting the
spectra directly from the original 2-dimensional frames for a subsample with large
Ly$\alpha$  extension.    In addition, a similar result was  also found
independently by  \cite{momose16}. Finally, we also observe a weak anti-correlation between Ly$\alpha$ EW and the UV size in agreement with \cite{law12}.
All the three parameters, therefore, seem to show some level of correlation.

So far we have compared  the flux density ratio of the different sub-samples and not their relative escape fraction of LyC photons, that depends  also on the intrinsic luminosity ratio and on the mean trasmissity of the sub-sample considered (see Equation \ref{eq:fesc}). We could reasonably conclude that galaxies in the significant sub-samples likely emit  more ionizing radiation, only  if the intrinsic luminosity ratio and the transmissivity do not intervene in changing our results. The transmissivity depends on the redshift of the sources (see Section \ref{sec:fesc}), and since  our sub-samples have very similar median redshifts, the transmissivity is approximately the same (within $10\%$) for all the sub-samples. The intrinsic luminosity ratio depends instead on different galaxies' properties, in particular the age of the stellar populations \cite[see for example Table 3 in][]{lucia}. As evaluated in \cite{grazian16}, for typical star-forming galaxies at $z\sim 3$, it varies between $1.7$ and $7.1$ for ages between $1\, Myr$ and $0.2\, Gyr$, adopting the \cite{bruzual} library. There is no evidence to date that galaxies which are  UV compact  or have a  Ly$\alpha$ with a small spatial extent,  have younger ages, so we expect that our sub-samples selected according to these two quantities have approximately the same age. For these sub-samples we can therefore interpret the differences in the measured density flux ratios in terms of relative escape fraction. On the other hand, galaxies with a very high Ly$\alpha$ EW are in general  believed to be younger than the rest of the star-forming galaxies population \citep[e.g.][]{malhotra02,gawiser07} and have therefore  lower intrinsic luminosity ratios. For this reason,  the  higher   flux density ratio  observed in the sub-sample with $EW(Ly\alpha)\geq70\,\AA$ with respect to its complementary, does not  necessarily have to be due entirely to a  higher relative escape fraction.
Since we have shown that the three parameters are marginally  correlated this could be also partially true for the other sub-samples:
however the correlation between  Ly$\alpha$ EW and the spatial extent (both UV and Ly$\alpha$) is quite weak. Finally, we note that the differences between the flux density ratios obtained for the interesting sub-samples and those obtained for the complementaries, are too large to be entirely erased by a change in the intrinsic luminosity ratio between the sub-samples.

Finally, we tested that the errors on the individual measurements (EW(Ly$\alpha$), Ly$\alpha_{ext}$ and $r_{UV}$) do not influence the results of the stacks, by  performing Monte Carlo simulations. For each parameter, we created 100 different versions of the original subset by varying each real measured value within a Gaussian distribution with the error on the measure as sigma. For each of these new subsets, we then re-selected the interesting sub-samples (e.g. $EW(Ly\alpha)\geq 70\AA$), bootstrapped them and evaluated the mean flux density ratio. For the Ly$\alpha$ EW and the UV effective radius, we obtained flux density ratio distributions that were completely inside the intervals shown in Figure \ref{fig:fratio}. For the Ly$\alpha$ spatial extent, we obtained a distribution that is slightly larger than the flux density ratio interval shown in Figure \ref{fig:fratio}, even though it is still much higher than the complementary sample. This test proves the validity of our sub-sample division, even considering the uncertainties on the parameters. 

The results obtained for the velocity shift of the Ly$\alpha$ with respect to the systemic redshift, are not in agreement with theoretical expectations \citep{dijkstra16,verhamme15}. 
This could be due to several reasons: first, our evaluation of the systemic redshift could be wrong, since it relies in most cases on an average relation that was tested only at lower redshift and the uncertainties on the measurements are relatively high. Second,  we know that in  $\sim30\%$ of the galaxies
the Ly$\alpha$ emission is actually double peaked \citep[e.g.][]{yamada12,kulas12}, but the peak separation is smaller than the VUDS resolution 
and we might be underestimating the velocity of the main (red) peak. Therefore
the division in sub-samples according to the velocity offset could be incorrect. 
Last, we cannot entirely discard the possibility that there is no real relation between the leakage of LyC emission and the velocity offset of the Ly$\alpha$ line, contrary to model predictions.

\section{Estimating $f_{esc}^{rel}$ after correction from contamination}
\label{sec:contamination}
       
As explained in Section \ref{sec:samplesel}, our sample is most probably contaminated in the sense that some of the objects selected could have nearby lower redshift faint galaxies (interlopers) that contribute to the flux in the LyC range of the extracted spectrum. 
For this reason, it would be incorrect to simply transform the values of the flux density ratio derived above into a LyC escape fraction. Since most of the galaxies in our sample ($\sim 165$ galaxies) do not have multi-band high resolution HST imaging, the careful cleaning procedure performed  in \cite{marchi} and \cite{lucia} is  not possible. In principle, we do not expect any correlation between contamination and the physical properties analysed in this work. If anything, we would expect a higher probability of contamination in objects with a more extended profile in the UV rest-frame than in compact objects, while the observations of the flux in the $895\AA$ region indicate the contrary if this was completely a contamination effect. In any case to transform the results obtained in the previous Section into relative escape fractions of LyC photons, we need to estimate the amount of flux that could come from the lower redshift interlopers in each sub-sample.  

One way to do this is to perform simulations, following the method outlined in \cite{vanzella10}.
The procedure is fully explained in the Appendix A: briefly  we used a very deep U-band image of the CDFS field \citep{nonino09}, which corresponds to the LyC flux rest-frame for sources at redshift $\sim 4$, to determine  the expected average integrated contribution of the foreground blue sources, by placing different rectangular apertures (corresponding to our spectroscopic slits) and performing Monte Carlo simulations. 
We find that the median value of the simulated flux, which corresponds to the contamination,  is very similar for all of our sub-samples, depending only slightly on the sample-size, in the sense that larger samples have slightly larger median contamination, thus validating our previous assumptions. On the other hand, the estimated contamination depends strongly on the maximum magnitude of possible contaminants that we set in the simulations (Umax, see Section \ref{sec:samplesel}). 

\begin{figure}
	\includegraphics[width=1.1\linewidth]{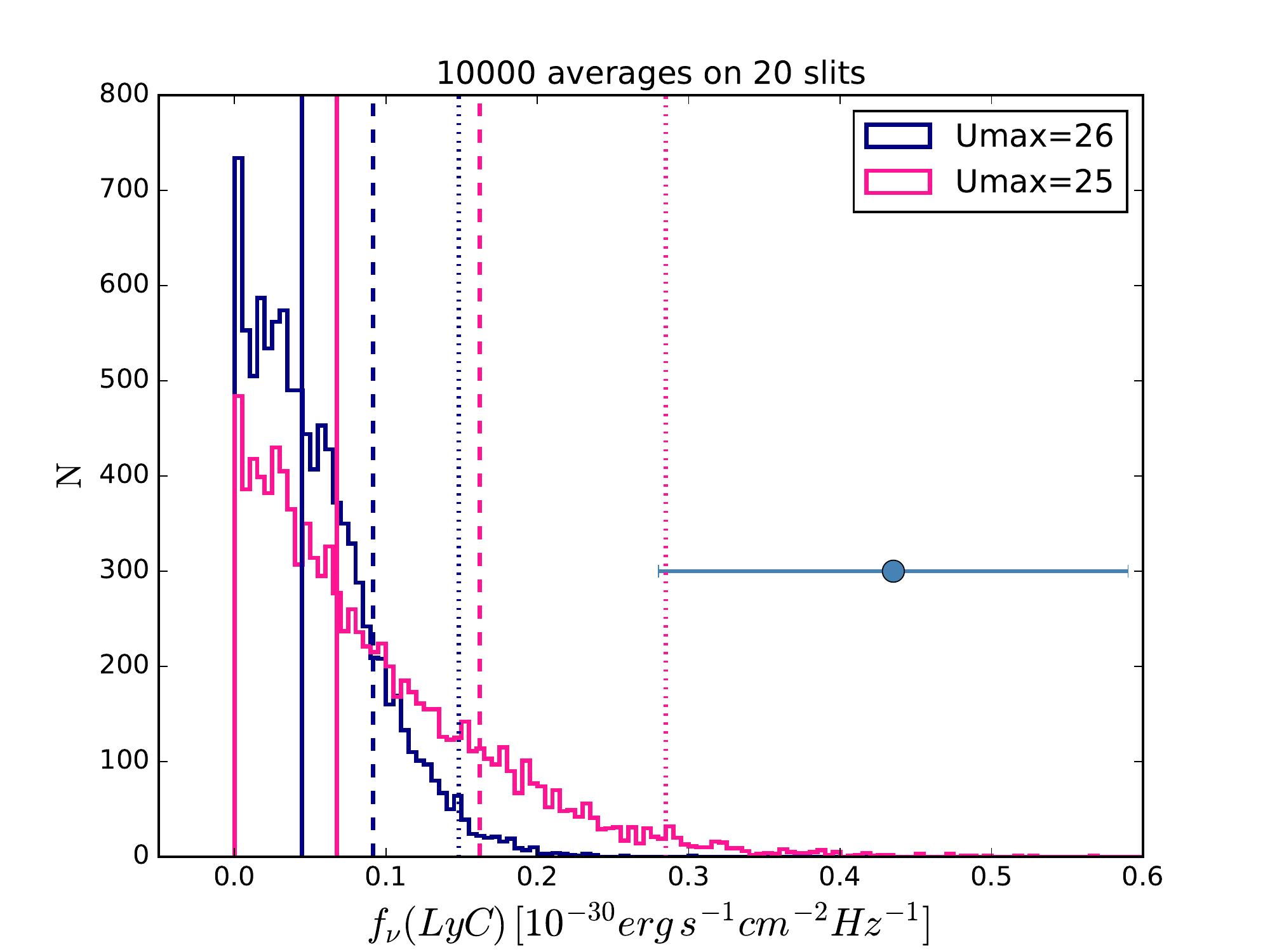}
	\caption{Distribution of the simulated observed-frame $f_{\nu}^{sim}$ coming from foreground sources for the sub-sample with $Ly\alpha_{ext}\le 5.7\,kpc$ imposing $U_{max}=25$ (\emph{magenta histogram}) and $U_{max}=26$ (\emph{blue histogram}).  The continuum vertical lines are the median of the distributions and the dashed and dotted lines are the $1\sigma$ and $2\sigma$ confidence intervals, respectively. The symbol on the Figure is the average observed-frame $f_{\nu}^{obs}(895)$ obtained for the sub-sample.}
	\label{fig:cont_LyAext}
\end{figure}

 Adopting a conservative approach and setting $U_{max}=25$, we proceed to estimate the relative escape fraction of LyC photons of the significant sub-samples, i.e. the samples where we believe there is real LyC signal in the $895\,\AA$ region. We therefore run  our simulations for the sub-samples selected as $EW(Ly\alpha)\ge 70\,\AA$, $Ly\alpha_{ext}\le 5.7\,kpc$ and $r_{UV}\le0.30\,kpc$ to obtain the contamination. As an example, we show in Figure \ref{fig:cont_LyAext} the distribution of the simulated fluxes for the sub-sample with $Ly\alpha_{ext}\le 5.7\,kpc$. We show both the distribution obtained assuming $U_{max}=25$ (magenta histogram) and that assuming  $U_{max}=26$ (blue histogram). In both cases the observed flux, which is the symbol in the Figure, is much higher than the simulated value, implying that there is probably real LyC escaping the galaxies in the sub-sample.

 To evaluate the average relative escape fraction, we first evaluated the \emph{real} LyC flux of our sub-samples, $f_{\nu}(LyC)$, subtracting the $2\sigma$ percentile of the distribution obtained with the simulations from the observed value, $f_{\nu}^{obs}(895)$,  and then we estimated the resulting flux density ratio, $\frac{f_{\nu}(LyC)}{f_{\nu}(1470)}$. 
To convert it to an escape fraction,  we evaluated the transmissivity averaging the individual galaxy's transmissivities of each sub-sample (0.25 for the $EW(Ly\alpha)\ge 70\,\AA$ sub-sample, 0.22 for the $Ly\alpha_{ext}\le 5.7\,kpc$ sub-sample and 0.28 for the  $r_{UV}\le 0.30\, kpc$ sub-sample). The individual values have been evaluated using  the analytical prescription given by \cite{inoue:transmissivity}. For easier comparison with earlier studies \citep[e.g.][]{steidel,grazian16,grazian17,marchi}, we adopted a value of $L_{\nu}(1470)/L_{\nu}(895)=3$, which corresponds to young star-forming galaxies with age $\sim 10$ Myr, assuming a constant Bruzual \& Charlot SFH and a Chabrier IMF \citep{chabrier03}. We point out that this is simply a
	multiplicative factor in the evaluation of the relative escape fraction,
	and it is therefore possible to re-scale$f_{esc}^{rel}$  with other values of
	$L_{\nu}(1470)/L_{\nu}(895)$ if needed.

We obtain upper limits of  $f_{esc}^{rel}\sim 33\%$ for the sub-sample with $EW(Ly\alpha)\ge 70\,\AA$ ($67\%$ if we subtract the $1\sigma$ percentile of the simulated distribution), $f_{esc}^{rel}\sim 48\%$ for the sub-sample with $Ly\alpha_{ext}\le 5.7\,kpc$ ($85\%$ subtracting  $1\sigma$) and $f_{esc}^{rel}\sim 23\%$ for the sub-sample with $r_{UV}\le 0.30\, kpc$ ($50\%$ subtracting  $1\sigma$).
Clearly these values are just indicative, since they crucially depend on the real contamination that is present in each  sub-sample.   Given the small sizes of these  significant sub-samples (14, 20 and 20 objects respectively) and the fact that there are several objects in common,  even the presence of a single contaminated source could sensibly change the results. In addition, as evident from Figure \ref{fig:cat_fin_cont} in the Appendix, with this choice of $U_{max}$ and slits size in the simulations, we are probably underestimating the contamination since the flux obtained from the simulations does not account for the flux in the full sample stack. Note however, that if instead of using the simulations to estimate the contamination, we simply assume that the full
sample of 201 galaxies is dominated  by contamination and subtract its flux density ratio  from the significant subsamples, we obtain similar high values for  $f_{esc}^{rel}$ for the three samples above.

The values we find for  $f_{esc}^{rel}$ are quite high and could seem at odd with observational results finding lower average values or very stringent upper limits \citep{boutsia,grazian16,lucia}: however the significant sub-samples represent only less than $10\%$ of the total galaxy population. If the galaxies with high  LyC emission are indeed  only a very small fraction of the entire  star forming population, we do not expect to detect significant emission when stacking large samples of objects without any pre-selection.

\section{Summary and conclusions}
In this work  we have  analysed some of the proposed correlations between LyC emission and other galaxies properties at high redshifts, exploiting the high quality and large number of spectra coming from the VUDS survey. We initially  selected  star-forming galaxies from the VUDS dataset at $3.5\leq z \leq 4.3$, paying particular attention in retaining only galaxies with a clean spectrum in the LyC region and galaxies with no clear contamination from bright neighbours in the same slit. For each of the 201 selected galaxies, we then evaluated (if possible)  the  Ly$\alpha$ EW, the Ly$\alpha$ velocity offset with respect to the systemic redshift, the Ly$\alpha$ spatial extension and the UV effective radius. Unfortunately, we could not estimate all these parameters for all the galaxies in the sample, because for example in some cases an evaluation of the systemic redshift was not possible, or the lack of high resolution HST data did not allow us to estimate the UV radius. To analyse the correlations between these parameters and LyC emission, we defined different sub-samples according to the properties predicted to	be good LyC emission indicators.  We therefore selected the most compact galaxies in Ly$\alpha$  ($Ly\alpha_{ext}\le 5.7\,kpc$) and UV ($r_{UV}\le0.30\,kpc$), the galaxies with the highest EW(Ly$\alpha$) ($EW(Ly\alpha)\ge 70\,\AA$) and those with the Ly$\alpha$ peak closest to the systemic redshift ($-200\le\Delta  v_{Ly\alpha}\le 200\,km/s$). Since we cannot reliably measure the flux in the LyC range for individual objects, we created spectral stacks of  all the galaxies in each sub-sample and estimated the average flux density ratios  ($\frac{f_{\nu}(895)}{f_{\nu}(1470)}$) from the stacks.
Our main results are the following:

 \begin{itemize}
\item[$\bullet$] We find that galaxies which are UV compact ($r_{UV}\le0.30\,kpc$) and have a high Ly$\alpha$ emission ($EW(Ly\alpha)\ge 70\,\AA$) are likely to have higher  
LyC emission, since these stack show a significant excess of flux in the LyC range compared to the complementary samples. This is  in agreement with theoretical studies \citep{dijkstra16,verhamme15,wise09} and with previous  observational studies at low redshift \citep{izotov16,verhamme17}. An indication that a high Ly$\alpha$ emission was related to the presence of LyC emission was also found in our recent work \citep{marchi}. 

\item[$\bullet$] We  find that  galaxies with a small Ly$\alpha$ spatial extent ($Ly\alpha_{ext}\le 5.7\,kpc$) have much larger fluxes in the LyC range, compared to the complementary samples. A possible relation between the spatial extent of Ly$\alpha$ and the presence of LyC emission has never been studied before.  According to our data, this parameter might even be 
more correlated to the presence of LyC emission than the other parameters, since the stack of the spatially small emitters shows a flux in the LyC range that is the highest of all the sub-samples.  

\item[$\bullet$] Although the above three parameters are correlated at some level, there are only very few objects where all three conditions ($r_{UV}\le0.30\,kpc$, $EW(Ly\alpha)\ge 70\,\AA$ and $Ly\alpha_{ext}\le 5.7\,kpc$) are met. Part of the reason is that  we could not measure all parameters for all galaxies in the initial sample.
These few objects are obviously the best candidates for being real LyC emitters: however since they are so few, their spectral stack is dominated by noise in the LyC range.


\item[$\bullet$]  We could  not apply any cleaning procedure to exclude contamination from lower-redshift interlopers in our samples, because multi-wavelength HST imaging, that is needed for identifying the contaminants, is not available for most of the sources. We believe  that the contamination should be approximately the same  in all the sub-samples, given that it should not correlate  with any of the galaxy properties. We 
attempted to estimate (and subtract) a  statistical contamination to the LyC flux using  Monte Carlo simulations performed on a  very deep U band image of the ECDFS field, which covers the observed wavelength of the LyC emission at $z\sim4$. We find that it is difficult  to give an accurate estimate of the real contamination  with the simulations because of the  uncertainties involved, especially in  the choice of the parameters to reproduce the observations ($U_{max}$, see Section \ref{sec:contaminatio}). Also, the small  number of galaxies in the significant sub-samples means that the simulated contamination flux has a very large distribution.  
For a reasonable choice of parameters in the simulation, we find that  comparing the observed   $f_{\nu}(895)f_{\nu}(1470)$ in the three significant sub-samples   to the simulated ones, there is a very high probability that  a significant fraction of the $f_{\nu}(895)$ flux comes from real LyC leakage, resulting in large escape fractions for the galaxies with very high $EW(Ly\alpha)$, small 
$r_{UV}$ and small $Ly\alpha_{ext}$.

\end{itemize}
The physical  picture that emerges from our results is that the
	conditions that regulate the escape of LyC and Ly$\alpha$ photons, must be
	closely related. Galaxies with compact UV  morphologies and compact and
	strong Ly$\alpha$ emission are the sources that most likely show LyC
	emission.
	This is in agreement with the scenario proposed by  \cite{nakajima14}
	where  compact star-forming
	regions can photoionise the ISM,  creating density-bounded regions or
	optically thin paths through the ISM. These
	paths could be seen as "holes" in the ISM caused for example by
	supernovae driven winds \citep{dove00,sharma16}, and would
	allow the simultaneous escape of LyC and Ly$\alpha$ radiation. The natural
	consequence would be the observed strong correlation between Ly$\alpha$ EW
	and LyC emission.
	In addition,   small HI column densities cause less scattering of the
	Ly$\alpha$ photons through the ISM,
	resulting in more compact Ly$\alpha$ spatial profiles \citep{yang16}. This would produce
	the observed anti-correlation between Ly$\alpha$ size and LyC
	emission.

        \begin{acknowledgements}
We thank the ESO staff for their continuous support for the VUDS survey,
particularly the Paranal staff conducting the observations and Marina Rejkuba and
the ESO user support group in Garching.
This work is supported by funding from the European Research Council Advanced Grant
ERC--2010--AdG--268107--EARLY and by INAF Grants PRIN 2010, PRIN 2012 and PICS 2013.

AC, OC, MT and VS acknowledge the grant MIUR PRIN 2010--2011.  
This work is based on data products made available at the CESAM data center,
Laboratoire d'Astrophysique de Marseille. 
 R.A. acknowledges support from the ERC Advanced Grant 695671 ‘QUENCH’.
 
 PC
 acknowledges support from CONICYT through the project FONDECYT regular
 1150216.
 
 AD is supported by the Polish National Science Centre grant UMO-2015/17/D/ST9/02121.
        \end{acknowledgements}
        
   \bibliographystyle{aa} 
   \bibliography{biblio.bib}
   
   \begin{appendix}
   	
   	\section{Statistical contamination from lower-redshift interlopers}
   	        \subsection{Method}
   	        \label{sec:contaminatio}
To simulate the contribution of low redshift interlopers to the flux in the LyC range we  follow the method outlined in \cite{vanzella10}.
   We start from a very deep U-band image of the ECDFS field \citep{nonino09}, which corresponds to the LyC flux rest-frame for sources at redshift $\sim 4$ as those in our sample, and we calculate the expected average integrated contribution of the foreground blue sources placing different
rectangular apertures (corresponding to our spectroscopic slits) and performing Monte Carlo simulations. Using the SExtractor algorithm \citep{sextractor}, we initially detected all sources down to a magnitude limit of about 30 and then generated a new U-band image composed only by the detected sources separated by null pixels requiring the SExtractor checkimage 'OBJECT'.
We did this to avoid the contamination from background fluctuations and therefore consider only the contamination from lower redshift interlopers. Since in the real case, relatively
bright neighbours are easily recognized as contaminants in the 2-dimensional spectra (that we have carefully checked), we have also
excluded from the image, the sources brighter than a
given U-mag ($U_{max})$ and replaced them with null pixels. In general it is also true that when planning spectroscopic observations of high redshift (faint)  sources,  slits would not be placed on targets  with very bright neighbours.
 
We then placed N random rectangular slits (with N the size of the sub-sample that we want to simulate) on the final image, and measured the mean flux coming from these slits. 
The rectangular apertures that we put on the image have dimensions $17\times 33\, pix^2$ which correspond  to $1\times 2\, arcsec^2$. $1\,arcsec$ mimics the slit width used to acquire the VUDS spectra and $2\,arcsec$ takes into account the fact that we cannot discriminate foreground sources closer than the seeing ($2\,\times \,seeing$). We  have assumed a mean seeing of $1\,arcsec$ which was the nominal condition of the VUDS survey, although observations were carried out in service mode and we do not have a precise statistics of the weather conditions during the acquisition of the spectra.

 Finally, we estimated the expected average
integrated contribution of the foreground sources  for each sub-sample, performing the above procedure 10000 times and evaluating the
median, 1$\sigma$  and 2$\sigma$  confidence levels of the mean $f_{\nu}^{sim}$ distribution. The median and dispersion of the final distribution  depend on the number N of slits and on the choice of $U_{max}$.
\subsection{Dependence on U$_{max}$}
\label{sec:cont_Umax}
The most delicate parameter to choose is $U_{max}$. Indeed, modifying this value,  significantly changes the results of our simulations. Following \cite{vanzella10}, we start by considering $U_{max}=25$ and $U_{max}=26$.

\begin{figure}
	\includegraphics[width=1.1\linewidth]{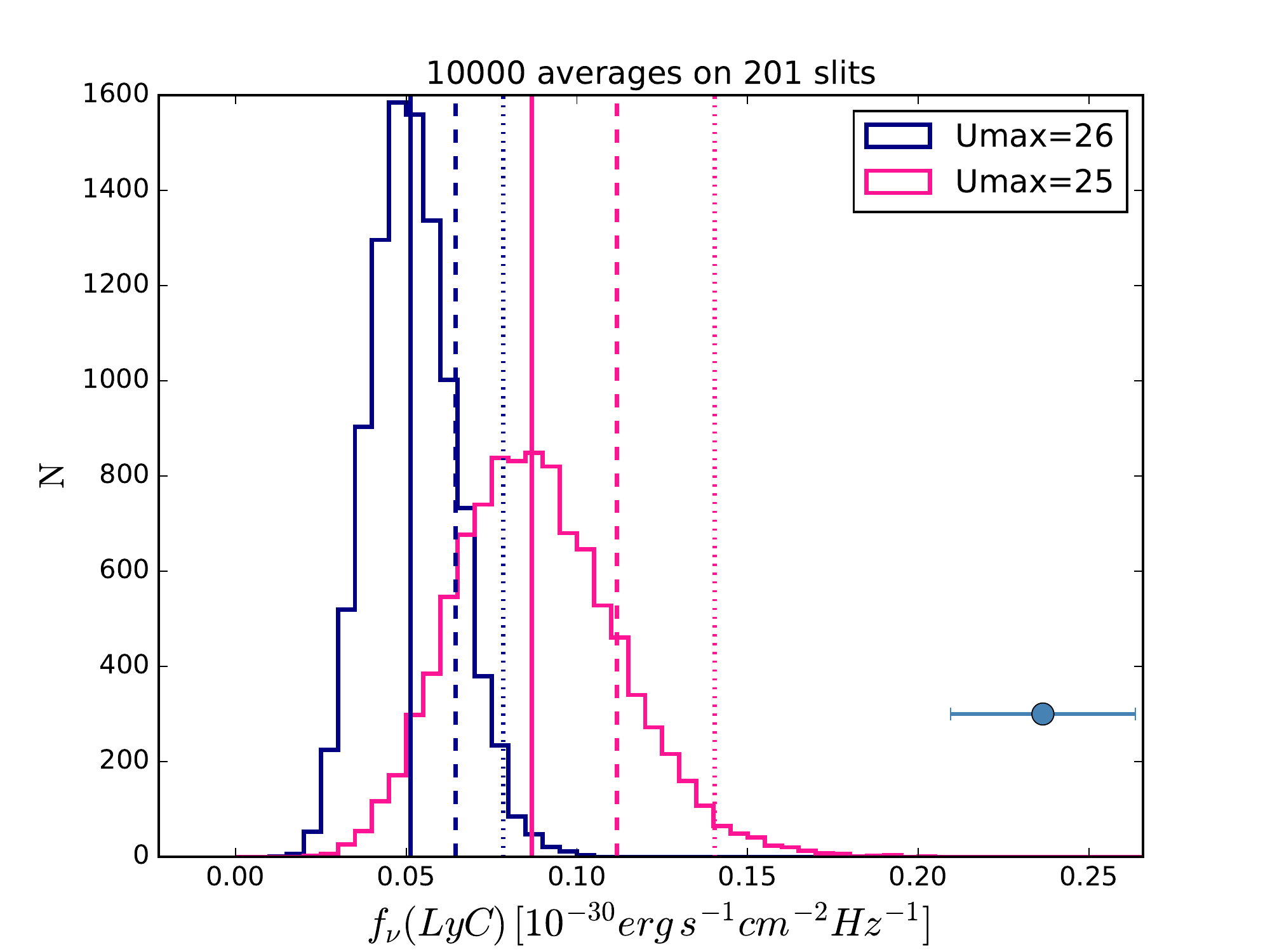}
	\caption{Distributions of the simulated observed-frame $f_{\nu}^{sim}$ coming from foreground sources for the total sample of 201 galaxies imposing $U_{max}=26$ (\emph{blue histogram}) and $U_{max}=25$ (\emph{magenta histogram}).  The continuum vertical lines are the medians of the two distributions and the dashed and dotted lines correspond to the $1\sigma$ and 2$\sigma$ confidence intervals. The symbol on the Figure is the average observed-frame $f_{\nu}^{obs}(895)$ obtained for the total sample.}
	\label{fig:cat_fin_cont}
\end{figure}  
In Figure \ref{fig:cat_fin_cont} we show the simulated observed-frame $f_{\nu}^{sim}$ coming from foreground sources for the total sample of 201 galaxies. The blue histogram (magenta histogram) is the distribution of the observed-frame $f_{\nu}^{sim}$ obtained imposing $U_{max}=26$ ($U_{max}=25$). The continuum vertical lines are the medians of the two distributions, and  the dashed and dotted lines correspond to the $1\sigma$ and $2\sigma$  confidence intervals respectively. The symbol on the Figure is the average observed-frame $f_{\nu}^{obs}(895)$ obtained for the total sample.
From the stack we cannot directly measure the average flux in the LyC range, because we normalise each spectrum at its value in the UV range during the stacking procedure.
	For this reason, we get $f_{\nu}^{obs}(895)$ multiplying the flux density ratio with the mean UV flux of the sample. To evaluate the latter, we averaged, for each galaxy in the sample, the signal in the UV range of the spectrum ($1420-1520\AA$) and then evaluated the mean and the $1\sigma$ error using the bootstrapping technique. 

This simple exercise outlines the effect of the choice of $U_{max}$ on the results of the simulations. Indeed, if we assume $U_{max}=25$, the $2\sigma$ value of the distribution of the simulated flux is about twice the value of the distribution obtained assuming $U_{max}=26$, implying a much higher contamination. In both cases the observed  $895\,\AA$ flux  is much higher than the simulated values. This could imply that the total sample contains not only contaminated sources but also  some real  LyC emitting galaxies, although this possibility is quite unlikely since we expect that the LyC emission is almost wiped out in the stack.  Most probably our simulations underestimate the real contamination, at least for this particular choice of parameters. One further possibility is that our choice of slits width is too optimistic and the spectra contain scattered light from objects that are located just outside the slit. We explore this scenario in Section \ref{sec:slitdim}.

\subsection{Dependence on the sample size}
\begin{figure}
	\includegraphics[width=1.1\linewidth]{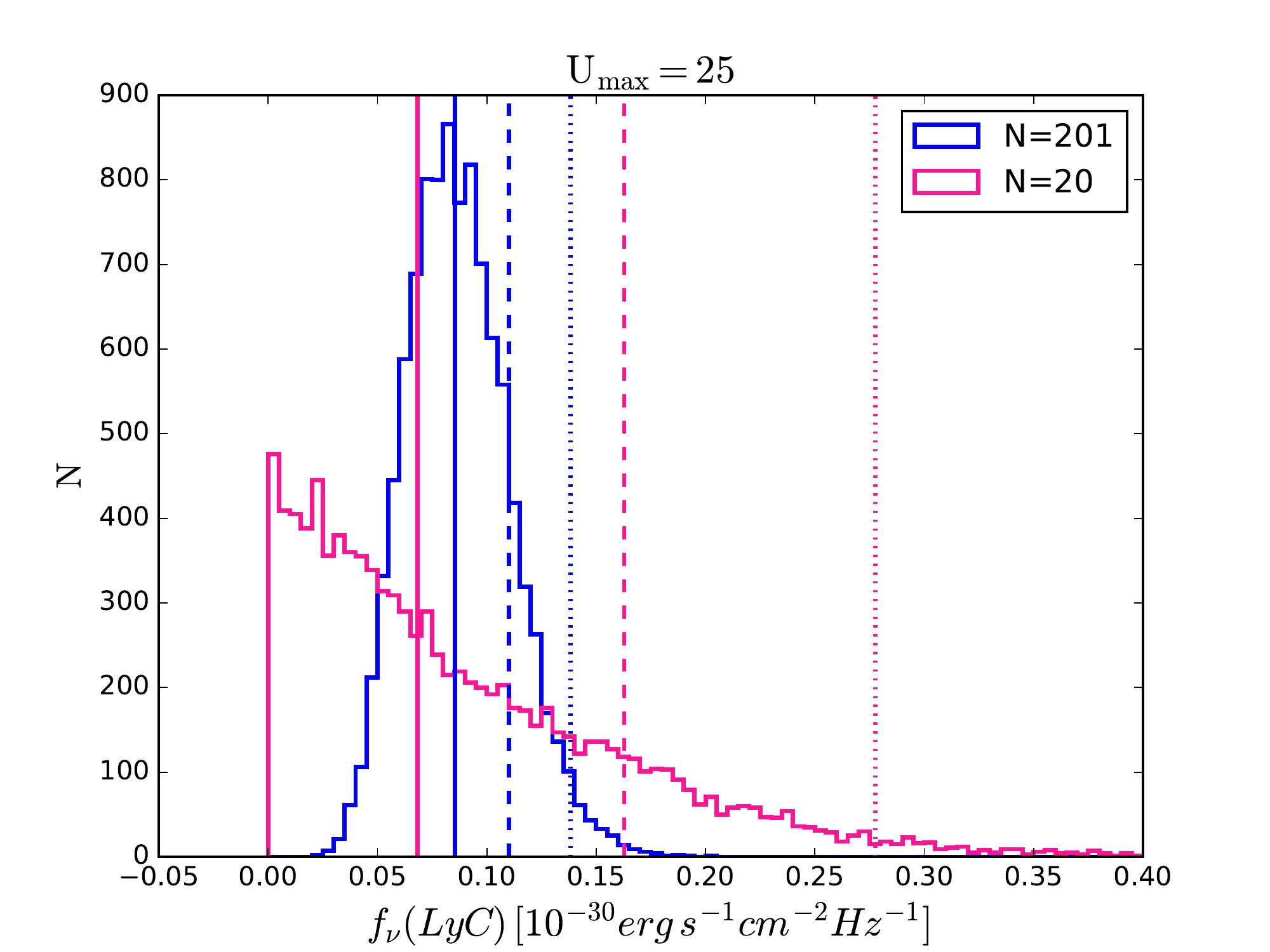}
	\caption{Comparison of the two distributions of the simulated observed-frame $f_{\nu}^{sim}$ coming from foreground sources for the total sample (\emph{blue histogram}, 201 slits per simulation) and for the sub-sample with $Ly\alpha_{ext}\leq5.7\,kpc$ (\emph{magenta histogram}, 20 slits per simulation) using $U_{max}=25$.  The continuum vertical lines are the medians of the two distributions, the dashed lines correspond to the $1\sigma$ confidence intervals while the dotted lines to the $2\sigma$ confidence intervals. }
	\label{fig:Umax25}
\end{figure}
  The shape of the distribution of the simulated flux, as well as the median and the $1\,\sigma$ and $2\,\sigma$ values,  depend also on the number of slits putted on the image in each simulation. We show in Figure \ref{fig:Umax25} the two distributions obtained for the total sample of 201 galaxies (blue histogram) and for the sub-sample with  $Ly\alpha_{ext}\leq5.7\,kpc$ (magenta histogram), which is formed by 20 galaxies, using the same $U_{max}$ (we chose the value of $U_{max}=25$). It is now possible to see the effect of using a different number of slits on the distribution of the simulated contamination flux. Indeed, when a lower number of slits is used, the distribution is much more extended (magenta histogram). Furthermore, while the median value of the distribution (indicated in the Figure with a vertical line) does not change much, only few percents as expected, the values of the $1\,\sigma$ (dashed lines) and $2\,\sigma$ (dotted lines) confidence intervals significantly change. 
  \begin{figure}
	\includegraphics[width=1\linewidth]{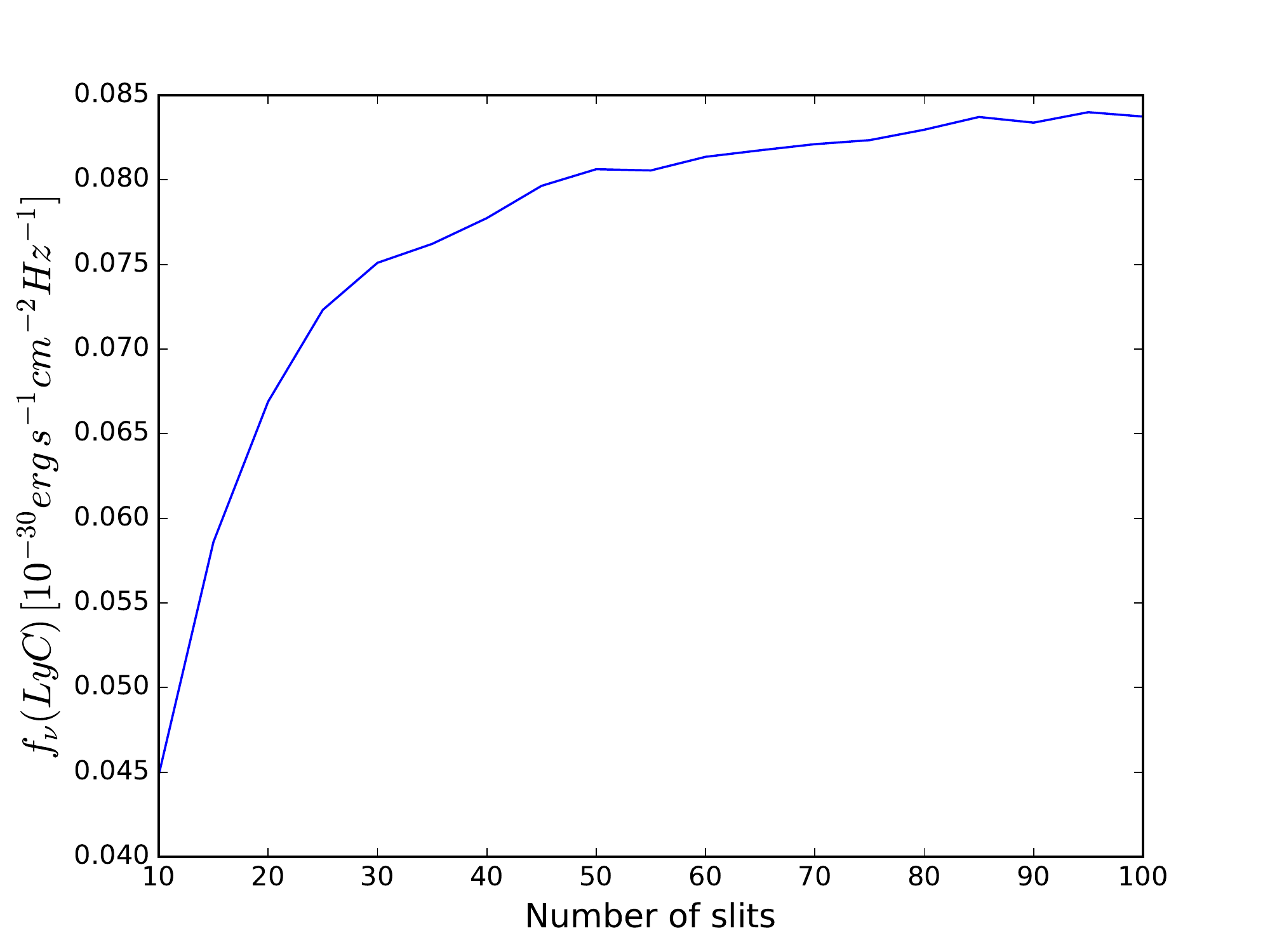}
  	\caption{Simulated contamination flux as a function of the number of slits used in the simulations. The flux in the Figure is the median value of the distribution and it is expressed in units of $10^{-30} erg\, s^{-1} cm^{-2} Hz^{-1}$.}
  	\label{fig:fsim_N}
  \end{figure}
  
  To validate the results obtained in Section \ref{results}, we need to verify the assumption that we made in Section \ref{sec:samplesel}, that is that all our sub-samples have approximately the same level of contamination. With this purpose, we performed a set of simulations for each value of N in a range between 10 and 100 with a step of 5, fixing $U_{max}=25$. We show in Figure \ref{fig:fsim_N} the median values of each distribution of simulated fluxes as a function of the number of slits used in the simulations. We obtain that the median contamination does not change much varying the number of slits used. Actually, Figure \ref{fig:fsim_N} shows that if we put in the simulations a smaller number of slits, the contamination is lower. Our most significant sub-samples are formed by very few galaxies compared to their complementaries (see Table \ref{tab:samples}), therefore the excess in the flux density ratio observed in these sub-samples cannot be interpreted as contaminated flux but as true emission. 
\subsection{Dependence on the slits dimension}
\label{sec:slitdim}
\begin{figure}
  	\includegraphics[width=1.1\linewidth]{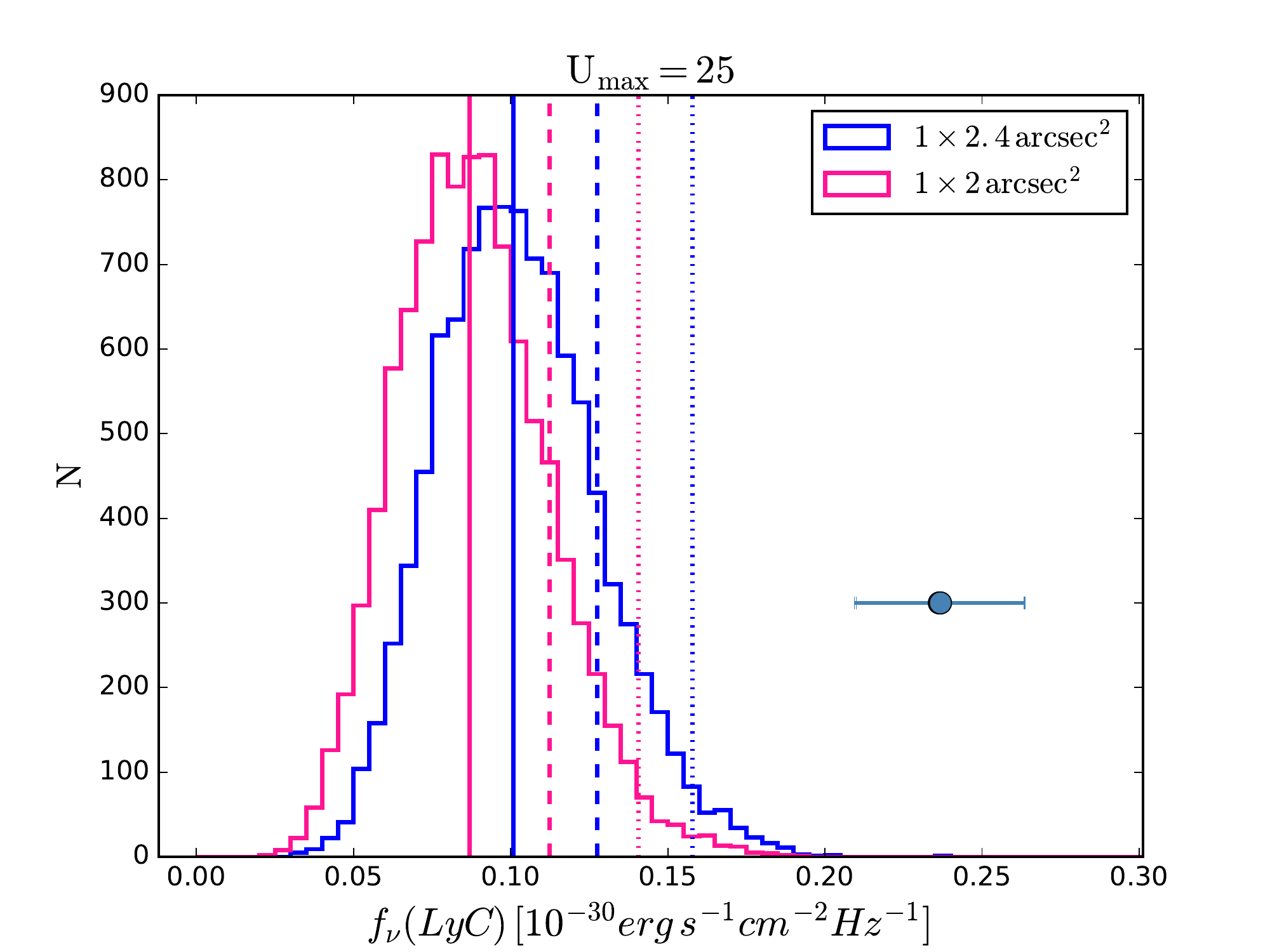}
  	\caption{Comparison of the two distributions of the simulated observed-frame $f_{\nu}^{sim}$ coming from foreground sources for the total sample (201 slits per simulation)  using  two different slits dimensions ($1\times2\, arcsec^2$, \emph{magenta histogram}, and $1\times2.4\, arcsec^2$, \emph{blue histogram}) and fixing $U_{max}=25$.  The continuum vertical lines are the medians of the two distributions, the dashed lines correspond to the $1\sigma$ confidence intervals while the dotted lines to the $2\sigma$ confidence intervals. The symbol on the Figure is the average observed-frame $f_{\nu}^{obs}(895)$ obtained for the total sample.}
  	\label{fig:fsim_slit}
  \end{figure}

As described in Section \ref{sec:contaminatio}, so far we have used slits of $1\times 2\, arcsec^2$, assuming an average seeing of $1\,arcsec$, which was the nominal condition of the VUDS survey. However, the observations were carried out by ESO in service mode so it is possible that the seeing was not always strictly within this limit.
In addition light could scatter into the slits from sources that are placed immediately outside the slit.
 For this reason we performed again our simulations using a slightly larger slit to test the dependence of the results. As an example, we use slits of $1\times2.4\, arcsec^2$ to simulate an average seeing of $1.2\,arcsec$ (the same effect would be obtained using larger slits to take into account scatter light).  In Figure \ref{fig:fsim_slit} we show the comparison of the two distributions of the simulated observed-frame $f_{\nu}^{sim}$ for the total sample (201 slits per simulation)  using  the different slit dimensions ($1\times2\, arcsec^2$, \emph{magenta histogram} and $1\times2.4\, arcsec^2$, \emph{blue histogram}) and fixing $U_{max}=25$. As expected, the distribution obtained using wider slits is shifted to higher values of contamination,  although the effect is lower than the one produced by the change in $U_{max}$ described in Section \ref{sec:cont_Umax}. Also in this case the observed  $895\,\AA$ flux value from the entire sample of 201 galaxies (which is the symbol in Figure \ref{fig:fsim_slit}) is larger than the simulated values for both slit sizes.

   \end{appendix}

\end{document}